\documentclass[lettersize,journal]{IEEEtran}

\usepackage{etoolbox}
\makeatletter
\patchcmd{\@makecaption}
  {\scshape}
  {}
  {}
  {}
\makeatletter
\patchcmd{\@makecaption}
  {\\}
  {.\ }
  {}
  {}
\makeatother

\usepackage{amsmath,amsfonts}
\usepackage{algorithmic}
\usepackage{algorithm}
\usepackage{array}
\usepackage[caption=false,font=normalsize,labelfont=sf,textfont=sf]{subfig}
\usepackage{textcomp}
\usepackage{stfloats}
\usepackage{url}
\usepackage{verbatim}
\usepackage{graphicx}
\usepackage{cite}
\usepackage{multirow}
\usepackage{booktabs}
\usepackage[colorlinks,linkcolor=blue,anchorcolor=blue,citecolor=blue]{hyperref}
\usepackage{xcolor}
\usepackage{forest}
\usepackage{adjustbox}
\usepackage{pifont} 
\usepackage{graphicx}
\usepackage{multirow}
\usepackage{amsmath}
\usepackage{amssymb}
\definecolor{light_blue}{rgb}{0.9137, 0.9647, 0.9961}
\definecolor{dark_blue}{rgb}{0.4588, 0.5804, 0.6902}
\definecolor{dark_green}{rgb}{0.0, 0.5, 0.0}
\definecolor{dark_red}{rgb}{0.785, 0.0, 0.0}
\definecolor{special_yellow}{rgb}{1, 0.7412, 0.0353}
\definecolor{special_green}{rgb}{0.4314, 0.6392, 0.596}
\newcommand{\tick}{\textcolor{special_green}{\ding{51}}}  
\newcommand{\cross}{\textcolor{special_yellow}{\ding{55}}}  

\begin{document}

\title{Generalist Virtual Agents: A Survey on Autonomous Agents Across Digital Platforms}

\author{
	
    Minghe Gao,
    Wendong Bu,
    Bingchen Miao,
    Yang Wu,
    Yunfei Li,
    Juncheng Li,

    Siliang Tang,
    Qi Wu,
    Yueting Zhuang,~\IEEEmembership{Senior Member,~IEEE},
    Meng Wang,~\IEEEmembership{Fellow,~IEEE}.
	\IEEEcompsocitemizethanks{
            \IEEEcompsocthanksitem Corresponding author: Juncheng Li.
		\IEEEcompsocthanksitem M. Gao, W. Bu, B. Miao, J. Li, S. Tang, and Y. Zhuang are with Zhejiang University, Hangzhou, China. E-mail: \{minghegao, wendongbu, miaobingchen23, junchengli, siliang, yzhuang\}@zju.edu.cn. 
		\IEEEcompsocthanksitem Y. Wu and Y. Li are with Antgroup, China. E-mail:\{wy306396, qixiu.lyf\}@antgroup.com
        \IEEEcompsocthanksitem Q. Wu is with the University of Adelaide, Adelaide, Australia. E-mail: qi.wu01@adelaide.edu.au
        \IEEEcompsocthanksitem M. Wang is with Hefei University of Technology, Hefei, China. E-mail: wangmeng@hfut.edu.cn
	}
}

\markboth{Journal of \LaTeX\ Class Files,~Vol.~X, No.~X, October~2024}%
{Shell \MakeLowercase{\textit{et al.}}: A Sample Article Using IEEEtran.cls for IEEE Journals}


\maketitle

\begin{abstract}
In this paper, we introduce the Generalist Virtual Agent (GVA), an autonomous entity engineered to function across diverse digital platforms and environments, assisting users by executing a variety of tasks. This survey delves into the evolution of GVAs, tracing their progress from early intelligent assistants to contemporary implementations that incorporate large-scale models. We explore both the philosophical underpinnings and practical foundations of GVAs, addressing their developmental challenges and the methodologies currently employed in their design and operation. By presenting a detailed taxonomy of GVA environments, tasks, and capabilities, this paper aims to bridge the theoretical and practical aspects of GVAs, concluding those that operate in environments closely mirroring the real world are more likely to demonstrate human-like intelligence. We discuss potential future directions for GVA research, highlighting the necessity for realistic evaluation metrics and the enhancement of long-sequence decision-making capabilities to advance the field toward more systematic or embodied applications. This work not only synthesizes the existing body of literature but also proposes frameworks for future investigations, contributing significantly to the ongoing development of intelligent systems.

\end{abstract}

\begin{IEEEkeywords}
Autonomous Agent, Intelligent Assistant, Implementation of Agent.
\end{IEEEkeywords}

\section{Introduction}

\begin{figure}
  \centering
    \includegraphics[scale=0.50]{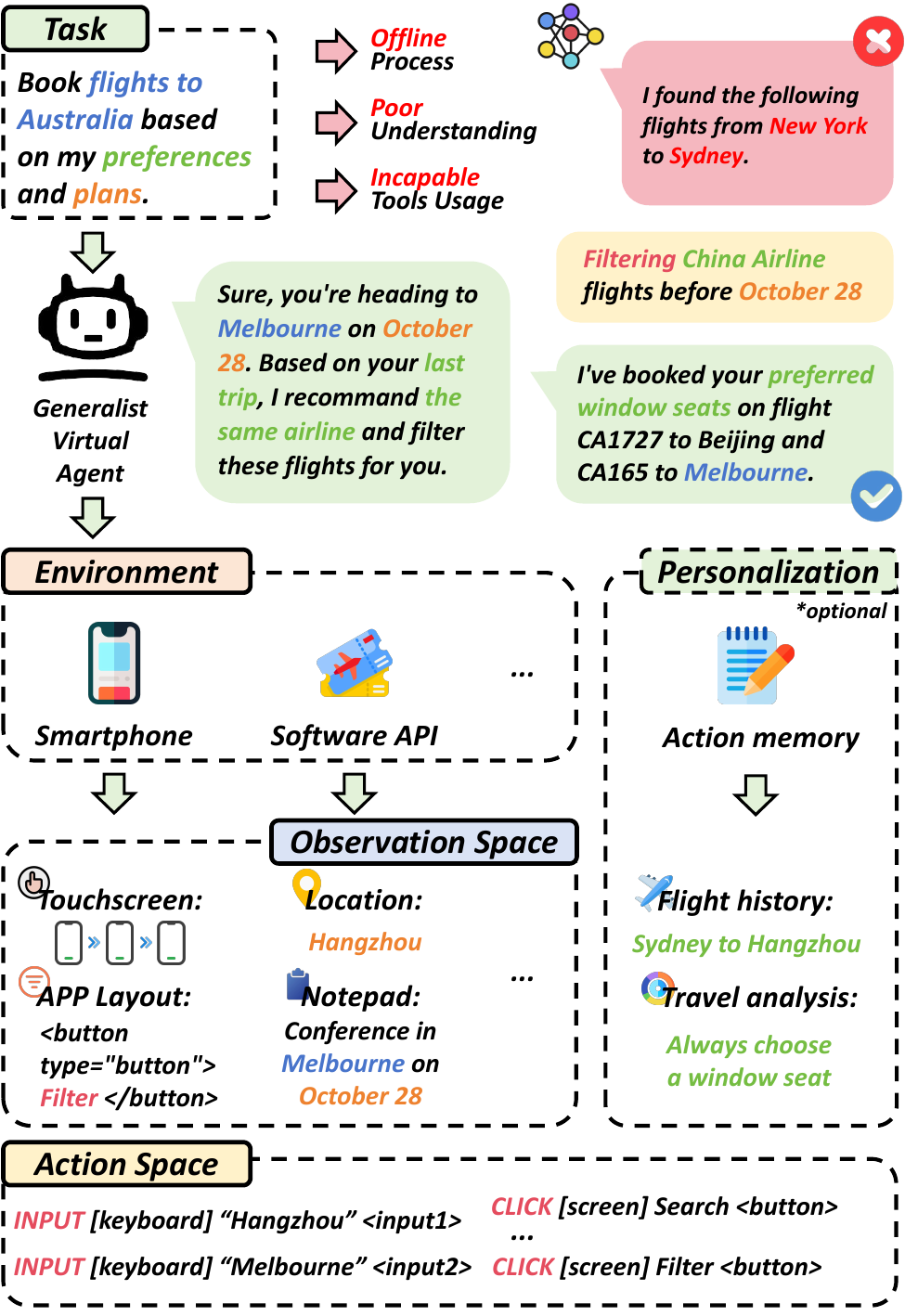}
     \vspace{-3mm}
    \caption{A comparison between MLLMs and GVAs: MLLMs demonstrate restricted aptitude in understanding intents and utilizing tools. Conversely, GVAs can systematically leverage resources and provide accurate responses.}
    \label{fig_1}
    \vspace{-4mm}
\end{figure}

\IEEEPARstart{T}he quest to develop a Generalist Virtual Agent (GVA) that approaches human-level intelligence in digital contexts underscores the evolution of Artificial Intelligence (AI). Initiated by the Turing Test~\cite{10.1093/mind/LIX.236.433}, research in AI aimed to create computational models capable of matching, and eventually surpassing human intelligence. Modern large-scale models~\cite{devlin2019bertpretrainingdeepbidirectional},~\cite{radford2018improving} are now endowed with sophisticated capabilities such as combinatorial reasoning~\cite{DBLP:conf/iccv/SurisMV23},~\cite{DBLP:conf/cvpr/GuptaK23},~\cite{DBLP:conf/emnlp/PanAWW23},~\cite{nsvqa} and tool usage~\cite{DBLP:conf/nips/SchickDDRLHZCS23},~\cite{wang2024aesopagentagentdrivenevolutionarystorytovideo}, making them prime candidates for GVAs. These goal-directed agents are distinguished from traditional computational tools by their ability to use these tools autonomously, offering personalized services and intelligent responses (Fig.~\ref{fig_1}). The pursuit of GVAs aims to create systems that can independently navigate various environments~\cite{DBLP:conf/iclr/MaLWHBJZFA24},~\cite{hong2023cogagentvisuallanguagemodel},~\cite{wang2024mobile},~\cite{zhang2024ufouifocusedagentwindows}, perform tasks, and interact with both users and other agents, representing a significant step to achieve artificial general intelligence.


The concept of an ``agent'' traces its philosophical roots back to thinkers such as Aristotle and Hume, encapsulating an entity’s capacity to harbor desires, beliefs, intentions, and to act. Transferred into computer science, agency denotes a process capable of engaging with other agents to collectively perform tasks on behalf of humans, often referred to as an ``intelligent assistant''. As AI evolves, the term ``agent'' comes to be central, describing \textbf{an entity that acts autonomously on a user’s behalf as software or computational models.}

\begin{figure*}
    \centering
    \begin{adjustbox}{max width=\textwidth, max height=\textheight}
        \begin{forest}
            for tree={
                grow=east,
                rounded corners,
                draw=dark_blue,
                parent anchor=east,
                child anchor=west,
                s sep=5pt, 
                anchor=west,
                edge={thick, draw},
                edge path={
                    \noexpand\path [draw, thick] (!u.east) -- ++(5pt,0) |- (.child anchor)\forestoption{edge label};}
            }
            [{\textbf{\textit{Generalist Virtual Agent}}}, minimum width=2cm, align=center, rotate=90, font=\LARGE
                [{\textbf{\textit{Future}}}, minimum width=4.7cm, align=center, edge path={\noexpand\path [draw, thick] (!u.south) -- ++(7.5pt,0) |- (.child anchor)\forestoption{edge label};}, font=\large
                    [{\textbf{\textit{From Virtual to Physical}} \S\ref{sec:7.2}}, minimum width=8cm, align=center, minimum height=0em
                        [{Xu et al.~\cite{xu2024survey}, Varley et al.~\cite{varley2024embodied}.}, text width=9.42cm, fill=light_blue]]
                    [{\textbf{\textit{From Individual to Systematic}} \S\ref{sec:7.1}}, minimum width=8cm, align=center, minimum height=0em
                        [{Shah et al.~\cite{shah2024multi}, Luzolo et al.~\cite{luzolo2024combining}, Han et al.~\cite{han2024llm}.}, text width=9.42cm, fill=light_blue]]]
                [{\textbf{\textit{Limitations}}}, minimum width=4.7cm, align=center, edge path={\noexpand\path [draw, thick] (!u.south) -- ++(7.5pt,0) |- (.child anchor)\forestoption{edge label};}, font=\large
                    [{\textbf{\textit{Heightened Security Concerns}} \S\ref{sec:6.4}}, minimum width=8cm, align=center, minimum height=0em
                        [{LLaVA-phi~\cite{zhu2024llava}, Smoothquant~\cite{xiao2023smoothquant}, AWQ~\cite{lin2024awq}.}, text width=9.42cm, fill=light_blue]]
                    [{\textbf{\textit{Limited Long-Sequence Decision-making}} \S\ref{sec:6.3}}, minimum width=8cm, align=center, minimum height=0em
                        [{MemoryBank~\cite{zhong2024memorybank}, Q*~\cite{wang2024q}, RAP~\cite{hao2023reasoning}.}, text width=9.42cm, fill=light_blue]]
                    [{\textbf{\textit{Insufficient Transferability}} \S\ref{sec:6.2}}, minimum width=8cm, align=center, minimum height=0em
                        [{SeeAct~\cite{zheng2024gpt4visiongeneralistwebagent}, Suglia et al.~\cite{suglia2024visually}.}, text width=9.42cm, fill=light_blue]]
                    [{\textbf{\textit{Unrealistic Environment and Dataset}} \S\ref{sec:6.1}}, minimum width=8cm, align=center, minimum height=0em
                        [{OSWorld~\cite{OSWorld}, Li et al.~\cite{li-etal-2020-mapping}, AppAgent~\cite{zhang2023appagentmultimodalagentssmartphone}.}, text width=9.42cm, fill=light_blue]]]
                [{\textbf{\textit{How to evaluate GVA?}}}, minimum width=4.7cm, align=center, edge path={\noexpand\path [draw, thick] (!u.south) -- ++(7.5pt,0) |- (.child anchor)\forestoption{edge label};}, font=\large
                    [{\textbf{\textit{MLLM-based}} \S\ref{sec:5.4}}, minimum width=4.01cm, align=center, minimum height=0em
                        [{VisualWebArena~\cite{koh2024visualwebarena}, WebVoyager~\cite{he2024webvoyager}, GUI-WORLD~\cite{chen2024guiworld}, OpenEQA~\cite{Majumdar_2024_CVPR}.}, text width=13.42cm, fill=light_blue]]
                    [{\textbf{\textit{Human}} \S\ref{sec:5.3}}, minimum width=4.01cm, align=center, minimum height=0em
                        [{WebVoyager~\cite{he2024webvoyager}, Li et al.~\cite{li-etal-2020-mapping}, AITW~\cite{NEURIPS2023_bbbb6308}, Mobile-Agent~\cite{wang2024mobile}.}, text width=13.42cm, fill=light_blue]]
                    [{\textbf{\textit{Detail}} \S\ref{sec:5.2}}, minimum width=4.01cm, align=center, minimum height=0em
                        [{\textit{Multi-dimensional}}, minimum width=4.02cm, align=center, inner ysep=0.3ex
                            [{ChatDev~\cite{chatdev}, MAC~\cite{tack2024onlineadaptationlanguagemodels}, Voyager~\cite{wang2023voyager}, GUI Odyssey\cite{lu2024gui}.}, text width=9cm, fill=light_blue]]
                        [{\textit{Set Inclusion}}, minimum width=4.02cm, align=center, inner ysep=0.3ex
                            [{WebShop~\cite{NEURIPS2022_82ad13ec}, META-GUI~\cite{sun-etal-2022-meta}, VisualWebArena~\cite{koh2024visualwebarena}.}, text width=9cm, fill=light_blue]]
                        [{\textit{Step-wise}}, minimum width=4.02cm, align=center, inner ysep=0.3ex
                            [{Li et al.~\cite{li-etal-2020-mapping}, WebSRC~\cite{chen-etal-2021-websrc}, WebVLN~\cite{chen2024webvln}, OmniACT\cite{kapoor2024omniactdatasetbenchmarkenabling}.}, text width=9cm, fill=light_blue]]]
                    [{\textbf{\textit{Overall}} \S\ref{sec:5.1}}, minimum width=4.01cm, align=center, minimum height=0em
                        [{RUSS~\cite{xu-etal-2021-grounding}, WebVoyager~\cite{he2024webvoyager}, Li et al.~\cite{li-etal-2020-mapping}, AITW~\cite{NEURIPS2023_bbbb6308}.}, text width=13.42cm, fill=light_blue]]]
                [{\textbf{\textit{How to implement GVA?}}}, minimum width=4.7cm, align=center, edge path={\noexpand\path [draw, thick] (!u.south) -- ++(7.5pt,0) |- (.child anchor)\forestoption{edge label};}, font=\large
                    [{\textit{\textbf{Strategy}} \S\ref{sec:4.3}}, minimum width=4.01cm, align=center, minimum height=0em
                        [{\textit{Cooperation or}\\\textit{Competition Strategy}}, minimum width=4.02cm, align=center, inner ysep=0.3ex
                            [{CMAT~\cite{liang2024cmat}, AUTOACT~\cite{qiao2024autoact}, ProAgent~\cite{zhang2024proagent}, ChatLLM~\cite{hao2023chatllm}, MAD~\cite{liang2023encouraging}, Multiagent Debate~\cite{du2023improving}, ChatEval~\cite{chan2023chateval}.}, text width=9cm, fill=light_blue]]
                        [{\textit{Reinforcement}\\\textit{Learning Strategy}}, minimum width=4.02cm, align=center, inner ysep=0.3ex
                            [{RL4VLM~\cite{zhai2024fine}, GLAINTEL~\cite{fereidouni2024search}, Juewu-mc~\cite{lin2021juewu}, Minerl 2020~\cite{guss2021towards}, PokeLLMon~\cite{hu2024pok}, SIMA~\cite{abi2024scaling}, LLM-X~\cite{lu2024mental}.}, text width=9cm, fill=light_blue]]
                        [{\textit{Fine-tuning Strategy}}, minimum width=4.02cm, align=center, inner ysep=0.3ex
                            [{COEVOL~\cite{li2024coevol}, Re-ReST~\cite{dou2024reflection}, ACT~\cite{chen2024learning}, Agent-FLAN~\cite{chen2024agent}.}, text width=9cm, fill=light_blue]]
                        [{\textit{Adaption Strategy}}, minimum width=4.02cm, align=center, inner ysep=0.3ex
                            [{ExpertPrompting~\cite{xu2023expertprompting}, Self-Refine~\cite{madaan2024self}, De-fine~\cite{gao2023fine}.}, text width=9cm, fill=light_blue]]]
                    [{\textit{\textbf{Model}} \S\ref{sec:4.2}}, minimum width=4.01cm, align=center, minimum height=0em
                        [{\textit{VLA-based Agent}}, minimum width=4.02cm, align=center, inner ysep=0.3ex
                            [{Rt-2~\cite{brohan2023rt}, Octo~\cite{team2024octo}, OpenVLA~\cite{kim2024openvla}, Lm-nav~\cite{shah2023lm}.}, text width=9cm, fill=light_blue]]
                        [{\textit{MLLM-based Agent}}, minimum width=4.02cm, align=center, inner ysep=0.3ex
                            [{AER~\cite{pan2024autonomous}, CLOVA~\cite{gao2024clova}, Agent Smith~\cite{gu2024agent}, CogAgent~\cite{hong2023cogagentvisuallanguagemodel}.}, text width=9cm, fill=light_blue]]
                        [{\textit{LLM-based Agent}}, minimum width=4.02cm, align=center, inner ysep=0.3ex
                            [{Agent-Pro~\cite{zhang2024agent}, AllTogether~\cite{liu2023alltogetherinvestigatingefficacyspliced}, CoELA~\cite{DBLP:conf/iclr/ZhangDSZDTSG24}.}, text width=9cm, fill=light_blue]]
                        [{\textit{Retriever-based Agent}}, minimum width=4.02cm, align=center, inner ysep=0.3ex
                            [{Glitho et al.~\cite{980543}, SAIRE~\cite{odubiyi1997saire}, ACQUIRE~\cite{das2002acquire}.}, text width=9cm, fill=light_blue]]]
                    [{\textit{\textbf{Environment}} \S\ref{sec:4.1}}, minimum width=4.01cm, align=center, minimum height=0em
                        [{\textit{Updated}}, minimum width=4.02cm, align=center, inner ysep=0.3ex
                            [{Mobile-Agent~\cite{wang2024mobile}, OSWorld~\cite{OSWorld}, AgentStudio~\cite{zheng2024agentstudiotoolkitbuildinggeneral}.}, text width=9cm, fill=light_blue]]
                        [{\textit{On-line}}, minimum width=4.02cm, align=center, inner ysep=0.3ex
                            [{WebShop~\cite{NEURIPS2022_82ad13ec}, WebArena~\cite{DBLP:conf/iclr/ZhouX0ZLSCOBF0N24}, VisualWebArena~\cite{koh2024visualwebarena}.}, text width=9cm, fill=light_blue]]
                        [{\textit{Off-line}}, minimum width=4.02cm, align=center, inner ysep=0.3ex
                            [{MiniWoB~\cite{ICLR17-Shi}, SeeClick~\cite{cheng2024seeclickharnessingguigrounding}, AITW~\cite{NEURIPS2023_bbbb6308}, MoTIF~\cite{burns2022motifvln}.}, text width=9cm, fill=light_blue]]]]
                [{\textbf{\textit{Why we need GVA?}}}, minimum width=4.7cm, align=center, edge path={\noexpand\path [draw, thick] (!u.south) -- ++(7.5pt,0) |- (.child anchor)\forestoption{edge label};}, font=\large
                    [{\textbf{\textit{Perspective of Application}} \S\ref{sec:3.3}}, minimum width=6cm, align=center, minimum height=0em
                        [{R-MAGIC~\cite{1316836}, RAP~\cite{kagaya2024rap}, AUTOACT~\cite{qiao2024autoact}, Plan4mc~\cite{baai2023plan4mc}, Reflexion~\cite{DBLP:conf/nips/ShinnCGNY23}.}, text width=11.41cm, fill=light_blue]]
                    [{\textbf{\textit{Perspective of HCI}} \S\ref{sec:3.2}}, minimum width=6cm, align=center, minimum height=0em
                        [{HyperPalm~\cite{DBLP:conf/smc/NazarovaBWFT22}, Auto-UI~\cite{zhang2024lookscreensmultimodalchainofaction}, Lin et al.~\cite{DBLP:conf/acl/LinFKD22}, Christiano et al.~\cite{DBLP:conf/nips/ChristianoLBMLA17}.}, text width=11.41cm, fill=light_blue]]
                    [{\textbf{\textit{Perspective of AI}} \S\ref{sec:3.1}}, minimum width=6cm, align=center, minimum height=0em
                        [{CogAgent~\cite{hong2023cogagentvisuallanguagemodel}, SeeClick~\cite{cheng2024seeclickharnessingguigrounding}, Pix2Act~\cite{DBLP:conf/nips/ShawJCBPHKLT23}, SeeAct~\cite{zheng2024gpt4visiongeneralistwebagent}, Auto-UI~\cite{zhang2024lookscreensmultimodalchainofaction}.}, text width=11.41cm, fill=light_blue]]]
                [{\textbf{\textit{What is GVA?}}}, minimum width=4.7cm, align=center, edge path={\noexpand\path [draw, thick] (!u.south) -- ++(7.5pt,0) |- (.child anchor)\forestoption{edge label};}, font=\large
                    [{\textit{\textbf{Action Space}} \S\ref{sec:2.4}}, minimum width=4.01cm, align=center, minimum height=0em
                        [{\textit{Others}}, minimum width=4.02cm, align=center, inner ysep=0.3ex
                            [{HyperPalm~\cite{DBLP:conf/smc/NazarovaBWFT22}.}, text width=9cm, fill=light_blue]]
                        [{\textit{Touchscreen}}, minimum width=4.02cm, align=center, inner ysep=0.3ex
                            [{AppAgent~\cite{zhang2023appagentmultimodalagentssmartphone}, AITW~\cite{NEURIPS2023_bbbb6308}, Auto-UI~\cite{zhang2024lookscreensmultimodalchainofaction}.}, text width=9cm, fill=light_blue]]
                        [{\textit{Mouse}}, minimum width=4.02cm, align=center, inner ysep=0.3ex
                            [{Pix2Act~\cite{DBLP:conf/nips/ShawJCBPHKLT23}, WebVoyager~\cite{he2024webvoyager}, WorkArena~\cite{workarena2024}.}, text width=9cm, fill=light_blue]]
                        [{\textit{Keyboard}}, minimum width=4.02cm, align=center, inner ysep=0.3ex
                            [{MiniWoB~\cite{ICLR17-Shi}, WebVoyager~\cite{he2024webvoyager}, WebArena~\cite{DBLP:conf/iclr/ZhouX0ZLSCOBF0N24}.}, text width=9cm, fill=light_blue]]]
                    [{\textit{\textbf{Observation Space}} \S\ref{sec:2.3}}, minimum width=4.01cm, align=center, minimum height=0em
                        [{\textit{Screen}}, minimum width=4.02cm, align=center, inner ysep=0.3ex
                            [{SeeClick~\cite{cheng2024seeclickharnessingguigrounding}, CogAgent~\cite{hong2023cogagentvisuallanguagemodel}, SeeAct~\cite{zheng2024gpt4visiongeneralistwebagent}, WebVLN~\cite{chen2024webvln}.}, text width=9cm, fill=light_blue]]
                        [{\textit{Document Object Model}}, minimum width=4.02cm, align=center, inner ysep=0.3ex
                            [{MiniWoB++~\cite{DBLP:conf/iclr/LiuGPSL18}, Mind2Web~\cite{NEURIPS2023_5950bf29}, CC-Net~\cite{DBLP:conf/icml/HumphreysRPTCMA22}.}, text width=9cm, fill=light_blue]]
                        [{\textit{Command Line Interface}}, minimum width=4.02cm, align=center, inner ysep=0.3ex
                            [{AgentBench~\cite{DBLP:conf/iclr/0036YZXLL0DMYZ024}, AgentStudio~\cite{zheng2024agentstudiotoolkitbuildinggeneral}, InterCode~\cite{DBLP:conf/nips/YangPNY23}.}, text width=9cm, fill=light_blue]]]
                    [{\textit{\textbf{Task}} \S\ref{sec:2.2}}, minimum width=4.01cm, align=center, minimum height=0em
                        [{\textit{Dialogue Task}}, minimum width=4.02cm, align=center, inner ysep=0.3ex
                            [{MAGIC~\cite{DBLP:conf/emnlp/ChenHMGPWFG21}, Zhao et al.~\cite{zhao2023chatgptequippedemotionaldialogue}.}, text width=9cm, fill=light_blue]]
                        [{\textit{Query Task}}, minimum width=4.02cm, align=center, inner ysep=0.3ex
                            [{Visual Programming~\cite{DBLP:conf/cvpr/GuptaK23}, LOGIC-LM~\cite{DBLP:conf/emnlp/PanAWW23}, HuggingGPT~\cite{DBLP:conf/nips/0001ST00Z23}.}, text width=9cm, fill=light_blue]]
                        [{\textit{Command Task}}, minimum width=4.02cm, align=center, inner ysep=0.3ex
                            [{AITW~\cite{NEURIPS2023_bbbb6308}, Pix2Act~\cite{DBLP:conf/nips/ShawJCBPHKLT23}, WebShop~\cite{NEURIPS2022_82ad13ec}, UFO~\cite{zhang2024ufouifocusedagentwindows}.}, text width=9cm, fill=light_blue]]] 
                    [{\textit{\textbf{Environment}} \S\ref{sec:2.1}}, minimum width=4.01cm, minimum height=0em, align=center
                        [{\textit{Operating System}}, minimum width=4.02cm, align=center, inner ysep=0.3ex
                            [{OSWorld~\cite{OSWorld}, AgentStudio~\cite{zheng2024agentstudiotoolkitbuildinggeneral}, MMAC-Copilot~\cite{song2024mmaccopilotmultimodalagentcollaboration}.}, text width=9cm, fill=light_blue]]
                        [{\textit{Application}}, minimum width=4.02cm, align=center, inner ysep=0.3ex
                            [{Mobile-Agent~\cite{wang2024mobile}, CogAgent~\cite{hong2023cogagentvisuallanguagemodel}, Auto-UI~\cite{zhang2024lookscreensmultimodalchainofaction}, MM-Navigator~\cite{yan2023gpt}, NS-VQA~\cite{nsvqa}, ViperGPT~\cite{DBLP:conf/iccv/SurisMV23}, Toolformer~\cite{DBLP:conf/nips/SchickDDRLHZCS23}.}, text width=9cm, fill=light_blue]]
                        [{\textit{Web}}, minimum width=4.02cm, align=center, inner ysep=0.3ex
                            [{MiniWoB~\cite{ICLR17-Shi}, WebArena~\cite{zhou2024webarena}, WebVoyager~\cite{he2024webvoyager}.}, text width=9cm, fill=light_blue]]]]
                [{\textbf{\textit{Introduction}}}, minimum width=4.7cm, align=center, edge path={\noexpand\path [draw, thick] (!u.south) -- ++(7.5pt,0) |- (.child anchor)\forestoption{edge label};}, font=\large]]
        \end{forest}    
    \end{adjustbox}
    \caption{The organization of the survey, with a curated list of representative works.}
\end{figure*}
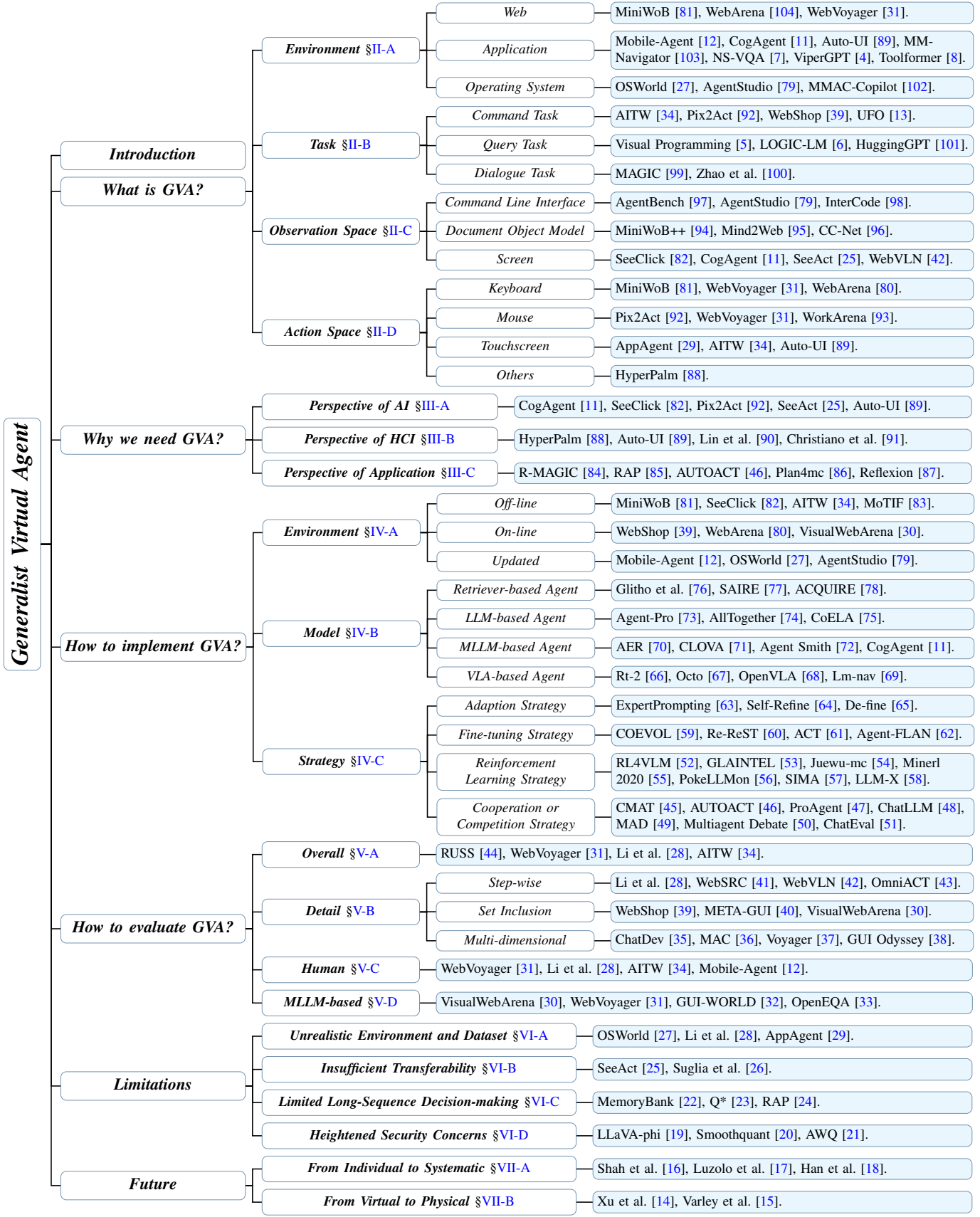

Since the 1950s, AI advancements have honed specific capabilities such as symbolic reasoning~\cite{chang2014symbolic}  and expertise in games like Go and chess~\cite{silver2018general}. However, research from that era confined agents to narrow, task-specific applications~\cite{NEURIPS2022_82ad13ec}, limiting them to broader scenarios. Despite introducing intelligent virtual assistants like Siri~\cite{siri_website} and Cortana~\cite{cortana_website}, these retrieval-driven and API-based systems have not yet achieved human-level intelligence. Designed to match voice inputs to API commands, they require frequent manual updates and patches as APIs and systems updated. This labor-intensive approach lacks genuine understanding. In contrast, an ideal agent should mirror humans' cognitive processes and interactive behaviors by using UI as the observation space and interacting through universal action spaces like keyboard, and screen interactions, serving as a powerful \textbf{Generalist Virtual Agent (GVA)}. Such an agent would excel in transferring and generalizing across multiple tasks and platforms, autonomously handling non-standard tasks by ``observing'' and ``acting''. This capability would especially benefit individuals with operational disabilities, enabling them to complete complex activities without reliance on predefined APIs.

\begin{figure*}[t]
\includegraphics[width=0.97\textwidth]{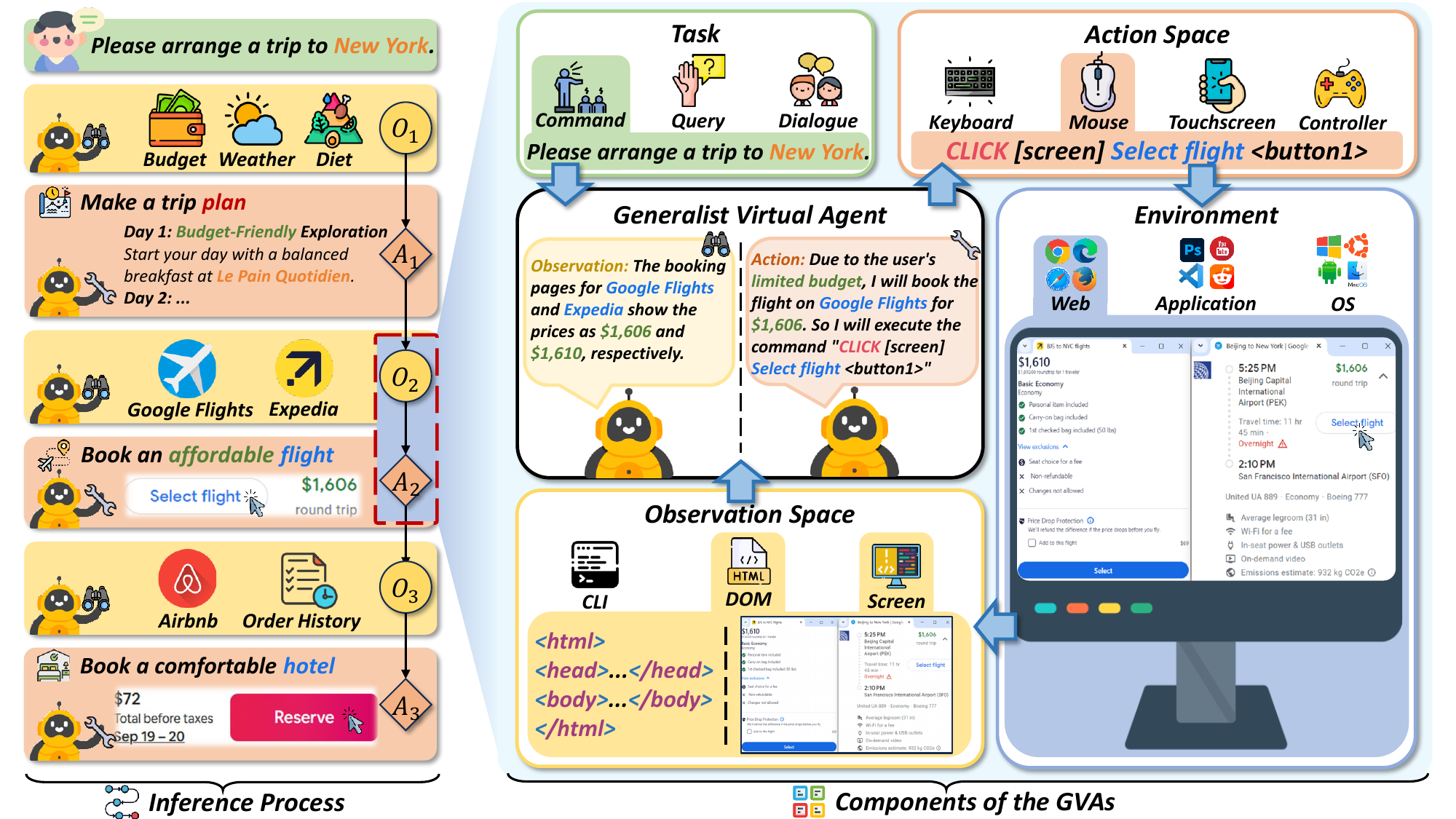}
\vspace{-3mm}
\centering\caption{An illustration of the GVA executing a travel arrangement task based on the user's command. The inference process (left) demonstrates the step-by-step decision-making of the GVA, incorporating user preferences to create a travel plan, select a flight, and reserve a hotel. The components of the GVAs (right) show the relationship between the task, action space, observation space, and environment.}
\label{fig_3}
\vspace{-3mm}
\end{figure*}

Encouragingly, the rise of large-scale models presents a promising avenue for agents to mimic human input-output methods, paving the way for GVAs. Agents powered by Large Language Models (LLMs) can now process structured texts like HTML~\cite{ICLR17-Shi},~\cite{DBLP:conf/iclr/LiuGPSL18},~\cite{NEURIPS2022_82ad13ec},~\cite{DBLP:conf/icml/HumphreysRPTCMA22} or images captions. Similarly, agents based on Visual Language Models (VLMs) can understand and localize images~\cite{DBLP:conf/nips/ShawJCBPHKLT23},~\cite{hong2023cogagentvisuallanguagemodel},~\cite{cheng2024seeclickharnessingguigrounding},~\cite{zhang2024lookscreensmultimodalchainofaction}. For instance, AutoDroid~\cite{DBLP:conf/mobicom/0004LLZYLJLZL24} combines the commonsense with app-specific knowledge through dynamic analysis. MM-Navigator~\cite{yan2023gpt} interacts with smartphone screens and determines subsequent actions to fulfill instructions. These advances highlight the growing research in agents, underscoring the need for a comprehensive survey to summarize current work and guide future research.

In response to the need, this paper provides a comprehensive investigation of GVA to address research gaps in this emerging field. We review prior studies and propose taxonomies and key principles for design and evaluation. Specifically, we explore four key questions: (1) \textit{What is a GVA?} (2) \textit{Why is a GVA necessary?} (3) \textit{How is a GVA implemented?} (4) \textit{What are the limitations and prospects of GVA?}


We define GVAs as systems operating in digital environments, performing tasks based on multimodal inputs, with their own observation and action spaces to mimic human logic (Section~\ref{sec:2}). Next, we highlight the urgent need for such agents from AI, HCI, and application perspectives (Section~\ref{sec:3}). Our literature review categorizes works by agent implementation, focusing on environments, model architectures, and learning strategies contingent upon task and data disparities(Section~\ref{sec:4}). For the emerging GVA technology lacking a comprehensive evaluation framework, we suggest using coarse- or fine-grained methods based on needs and exploring manual and model-based evaluations for qualitative assessment (Section~\ref{sec:5}). Finally, we discuss limitations, future directions, and insights for advancing GVA research (Section~\ref{sec:6} \&~\ref{sec:7}).

The contribution of this survey lies in its systematic examination of existing agent research, leading to the conclusion that GVAs closer to real-world environments are more likely to exhibit human-like intelligence. However, we express concerns that current GVAs overly rely on large-scale models. Should the development of these models reach a plateau, GVAs will need to explore alternative pathways. Thus, this survey provides new insights: proposing the evolution of tools into agent systems for human-computer interaction, or the breakthrough into embodied intelligence beyond digital confines.

\section{What is a generalist virtual agent?}
\label{sec:2}

In the preceding discussion, we briefly introduced GVA, an autonomous intelligent assistant designed to fulfill user needs. Previously, Google's Generalist Agent~\cite{reed2022a} is a neural network demonstrating diverse capabilities, from manipulating robot arms to image captioning. IBM's Virtual Agent~\cite{IBM_virtual_agent} integrates Natural Language Processing (NLP), intelligent search, and robotic process automation. Weng from OpenAI describes an ``Agent"~\cite{weng2023agent} as a universal problem solver that leverages LLMs for planning, memory, and tool use. However, these definitions exhibit limitations respectively: 1) insufficient consideration of the actual digital device environments, 2) neglected generalization across generic tasks, and 3) inadequate capabilities for multimodal perception.

To supplement existing definitions, we propose \textbf{Generalist Virtual Agent}, designed to utilize multimodal data for autonomous decision-making and task completion in a virtual setting. Furthermore, we summarize a GVA consisting of four integral parts: 1) \textbf{\textit{Environment}}: A virtual workspace for task execution (Section~\ref{sec:2.1}). 2) \textbf{\textit{Task}}: A set of instructions or queries for the GVA to fulfill (Section~\ref{sec:2.2}). 3) \textbf{\textit{Observation Space}}: Representation of information for understanding (Section~\ref{sec:2.3}). 4) \textbf{\textit{Action Space}}: Tools for environmental interaction and modification (Section~\ref{sec:2.4}). We illustrate the inference process and components of a GVA in Fig.~\ref{fig_3} and summarize some existing works in Table~\ref{table_1}.

\subsection{\textbf{Environment}}
\label{sec:2.1}

We define the GVA’s operational context as a virtual environment varying from simple web interactions to complex cross-application tasks with probable hardware involvement. Because current research lacks a unified classification of these environments, prompting our categorization based on GVA's runtime: 1) \textbf{\textit{Web}}: An environment supporting web-related tasks under browser constraints (Section~\ref{sec:2.1.1}). 2) \textbf{\textit{Application}}: An environment providing APIs and requirements for domain-specific knowledge (Section~\ref{sec:2.1.2}). 3) \textbf{\textit{Operating System}}: An environment managing hardware resources and software execution. (Section~\ref{sec:2.1.3}). This classification aims to cover the breadth of environments discussed in contemporary studies. We will next expound on these categories individually.

\subsubsection{\textbf{Web}}
\label{sec:2.1.1}

As an environment executing web-related tasks within browsers, the web environment is characterized by a relatively standardized content structure, notably HTML and XML, and supports fundamental operations such as confirm and return. This uniform structure equips the GVA with essential skills for efficient web navigation.

Early web environments used simplified or simulated settings, like MiniWoB~\cite{ICLR17-Shi} and MiniWoB++~\cite{DBLP:conf/iclr/LiuGPSL18} reduced the complexity of HTML, CSS, and JavaScript, diverging from real-world web. Later, projects such as WebShop~\cite{NEURIPS2022_82ad13ec}, WebArena~\cite{DBLP:conf/iclr/ZhouX0ZLSCOBF0N24}, and VisualWebArena~\cite{koh2024visualwebarena} simulated more realistic websites but excluded real-world challenges like captchas and pop-ups. Recent efforts like WorkArena~\cite{workarena2024}, Mind2Web~\cite{NEURIPS2023_5950bf29}, and WebVLN~\cite{chen2024webvln} pivot to using real websites, providing more authentic tasks for testing GVAs.

The diverse tasks in the web train GVA by enhancing multitasking skills and core execution abilities, make the web an ideal platform for developing GVA's operational proficiency.

\subsubsection{\textbf{Application}}
\label{sec:2.1.2}

The application environment is crucial for training GVAs in domain-specific tasks like image editing and code generation. Here, GVAs often incorporate expert external knowledge and utilize specialized applications to execute professional tasks effectively. This environment enhances the GVA's expertise in specific fields, fostering deeper understanding and greater proficiency in their specialized functions.

In the application environment, GVAs use predictive actions and APIs to execute tasks across diverse applications. Agents like AppAgent~\cite{zhang2023appagentmultimodalagentssmartphone} and Mobile-Agent~\cite{wang2024mobile} manage 10 applications each while MoTIF~\cite{burns2022motifvln} handles 125, including YouTube, TikTok, and Google Maps. AITW~\cite{NEURIPS2023_bbbb6308} offers a benchmark with 159 applications, 198 websites, and 715,000 demonstration segments. LOGIC-LM~\cite{DBLP:conf/emnlp/PanAWW23} handles logical reasoning by converting problems into symbolic representations. NS-VQA~\cite{nsvqa} combines a scene parser, question parser, and program executor to handle visual questions. ViperGPT and Visual Programming~\cite{DBLP:conf/cvpr/GuptaK23} calling visual expert models through the code. Toolformer~\cite{DBLP:conf/nips/SchickDDRLHZCS23} trains LLMs to autonomously use various tools, while AesopAgent~\cite{wang2024aesopagentagentdrivenevolutionarystorytovideo} creates videos from storylines using specialized tools.

Training in an application environment where tasks are distinctly specialized intensively develops GVAs' domain-specific expertise. This approach ensures that GVAs achieve a high degree of expertise in their respective fields.

\subsubsection{\textbf{Operating System}}
\label{sec:2.1.3}

The OS environment, which manages hardware resources and executes software, grants GVA broader permissions. This enables GVA to operate in a developer environment, switch between applications, and facilitate cross-application collaboration, expanding the range of tasks it can solve and enhancing its training as a generalist agent.

OSWORLD~\cite{OSWorld} and AgentStudio~\cite{zheng2024agentstudiotoolkitbuildinggeneral} offer environments and benchmarks for agent-OS interaction. MMAC-Copilot~\cite{song2024mmaccopilotmultimodalagentcollaboration} provides a team collaboration framework where agents assume roles like planner, librarian, programmer, or viewer, enhancing their OS interaction capabilities through a collaborative chain. UFO~\cite{zhang2024ufouifocusedagentwindows} utilizes the Microsoft UI Automation framework on Windows, greatly improving the agent's capabilities to interact with GUIs.

In the OS environment, GVAs encounter tasks from both web and application domains, offering a varied training landscape that enhances their ability to generalize across environments. This broad exposure makes the OS environment pivotal for developing adaptable and proficient GVAs.

\subsection{\textbf{Task}}
\label{sec:2.2}

\begin{figure}[t]
\includegraphics[width=\linewidth]{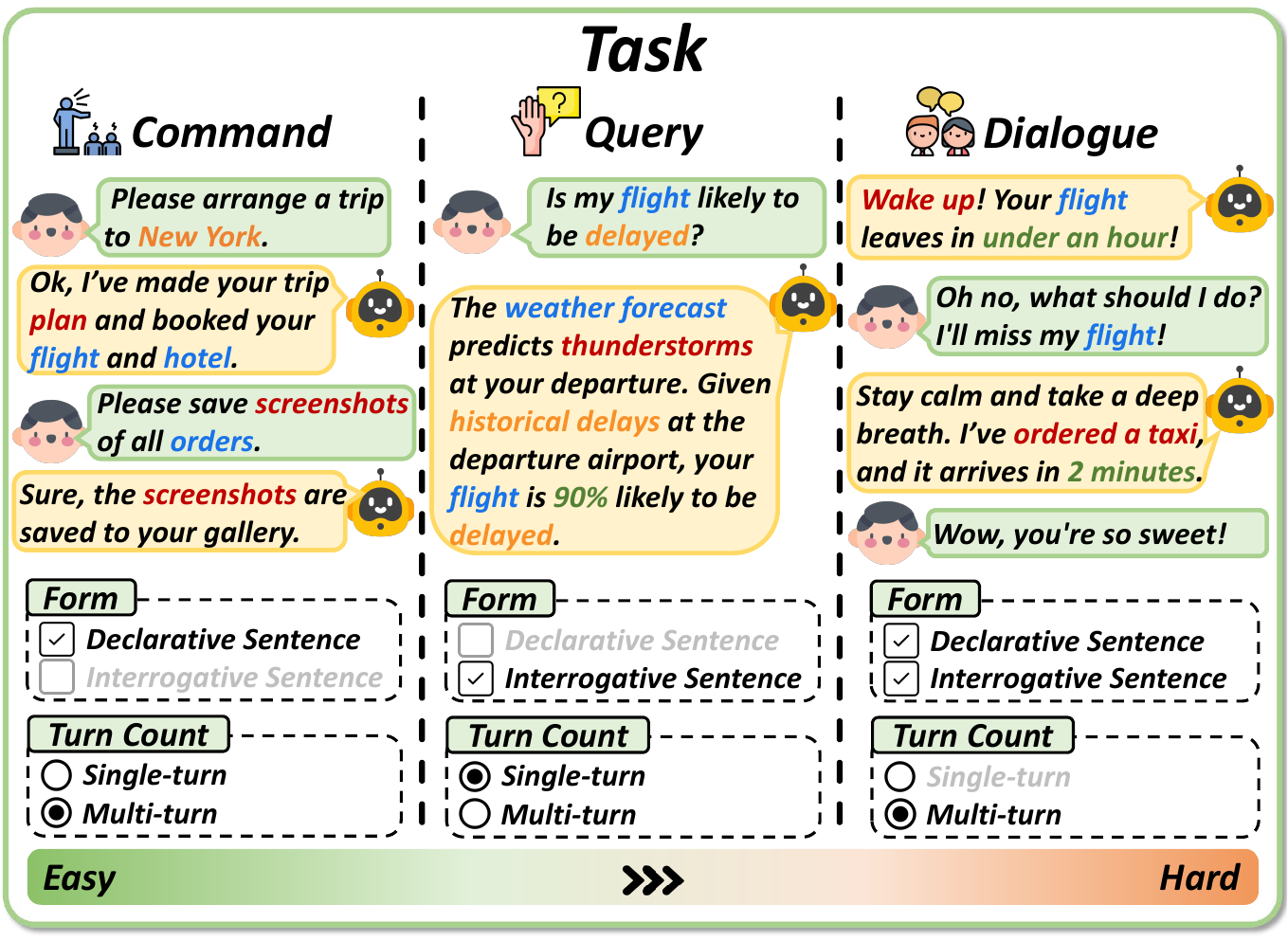}
\vspace{-8mm}
\centering\caption{Based on description clarity, we classify tasks from easy to difficult as follows: (1) \textbf{\textit{Command Task}}: Clear and declarative task description; (2) \textbf{\textit{Query Task}}: Interrogative tasks without step guidance; (3) \textbf{\textit{Dialogue Task}}: Multi-turn conversations with vague intentions.}
\label{fig_4}
\vspace{-2mm}
\end{figure}

In light of the diverse tasks in current research, we categorize GVA tasks based on description clarity, crucial for effective training, as shown in Fig.~\ref{fig_4}. Categories include: (1) \textbf{\textit{Command Task}}: Clear, declarative task description (Section~\ref{sec:2.2.1}); (2) \textbf{\textit{Query Task}}: Interrogative tasks without step guidance (Section~\ref{sec:2.2.2}); (3) \textbf{\textit{Dialogue Task}}: Multi-turn conversations with vague intentions (Section~\ref{sec:2.2.3}). 
Tasks guide the development of training strategies tailored to each task type, enhancing GVA performance across varied interactions.

\subsubsection{\textbf{Command Task}}
\label{sec:2.2.1}

We further classify command tasks for GVAs based on input form into three categories: \textit{a) Single Simple Command}: tasks with concise instructions; \textit{b) Single Detailed Command}: tasks that provide extensive details for execution; \textit{c) Multiple Progressive Commands}: a series of sequential commands requiring step-by-step completion.  

\textit{a) Single Simple Command}: This command type handles straightforward instructions. In AppAgent~\cite{zhang2023appagentmultimodalagentssmartphone} and AITW~\cite{NEURIPS2023_bbbb6308}, agents receive brief commands to perform tasks like navigation, messaging, or image editing. The simplicity of these task instructions allows for large-scale data collection, enhancing the GVA's basic capabilities.

\textit{b) Single Detailed Command}: This command type includes fine-grained guidance. Platforms like UFO~\cite{zhang2024ufouifocusedagentwindows}, Pix2Act~\cite{DBLP:conf/nips/ShawJCBPHKLT23} and WebShop~\cite{NEURIPS2022_82ad13ec} provide detailed instructions for complex tasks, such as multi-application coordination or targeted online shopping. These tasks, rich in detail, are ideal for enhancing the GVA's decision-making capabilities.

\textit{c) Multiple Progressive Commands}: Sequential tasks, such as route planning through progressive instructions, require GVAs to reference past interactions, fostering advanced contextual understanding. Each task strategically advances GVA competencies, tailoring challenges to developmental needs.

Whether simple, detailed, or progressive, command tasks have more rich accessible data and clearer task descriptions, facilitating the construction of large-scale instruction datasets and improving the GVA's instruction-following capabilities.

\subsubsection{\textbf{Query Task}}
\label{sec:2.2.2}

Query tasks for GVAs are characterized by their use of interrogative sentences leading to a relatively unclear task description and the requirement for GVAs to independently explore or query information. Based on the number of interactions, these tasks are divided into: \textit{a) Single Round Query} and \textit{b) Multiple Round Query}.

\textit{a) Single Round Query}: These tasks require a single interaction between the user and the GVA. Hence, we can regard them as VQA tasks. For example, ViperGPT~\cite{DBLP:conf/iccv/SurisMV23} and Visual Programming~\cite{DBLP:conf/cvpr/GuptaK23} use coding to answer visual questions. In LOGIC-LM~\cite{DBLP:conf/emnlp/PanAWW23}, it needs to tackle logical reasoning. HuggingGPT~\cite{DBLP:conf/nips/0001ST00Z23} and MM-REACT~\cite{yang2023mm} call on expert models for cross-modal problems, and WebVLN~\cite{chen2024webvln} include a VQA-based task designed for GVAs. These tasks demand that GVAs comprehend questions thoroughly and deliver satisfactory answers in a single exchange.

\textit{b) Multiple Round Query}: These tasks demand continuous dialogue, necessitating GVAs with architectures that support large context windows to retain information across multiple interactions. This setup ensures effective handling of complex, evolving queries over extended conversations.

As we conclude, query tasks are characterized by relatively unclear descriptions and the need for autonomous information gathering. These two characteristics contribute to training the GVA's intent detection and knowledge retrieval capabilities, promoting the GVA to analyze questions and generate answers.

\subsubsection{\textbf{Dialogue Task}}
\label{sec:2.2.3}

Unlike tasks requiring specific answers, dialogue tasks emphasize multi-turn, human-like interactions. These involve humor, contextual understanding, intent detection, linguistic naturalness, and user satisfaction, reflecting the complexity of human communication.

Microsoft's Copilot+PC~\cite{Microsoft_Copilot_Plus} recall function retrieves previously displayed content via conversation. At WWDC24, Apple launched Apple Intelligence~\cite{Apple_Intelligence}, which leverages personal context from apps like Mail and Calendar. MAGIC~\cite{DBLP:conf/emnlp/ChenHMGPWFG21} generates more human-like image comments, while Zhao et al.~\cite{zhao2023chatgptequippedemotionaldialogue} have shown promising results in studying ChatGPT’s capabilities for generating emotional dialogue.

Dialogue tasks require GVA to go beyond basic QA, simulating human thinking and communication for intelligent interaction. However, the lack of quantifiable evaluation metrics limits progress in this area. As a result, existing work falls short of realizing the full potential of dialogue tasks, leaving further exploration for future research.

\subsection{\textbf{Observation Space}}
\label{sec:2.3}
\begin{figure}[t]
\includegraphics[width=\linewidth]{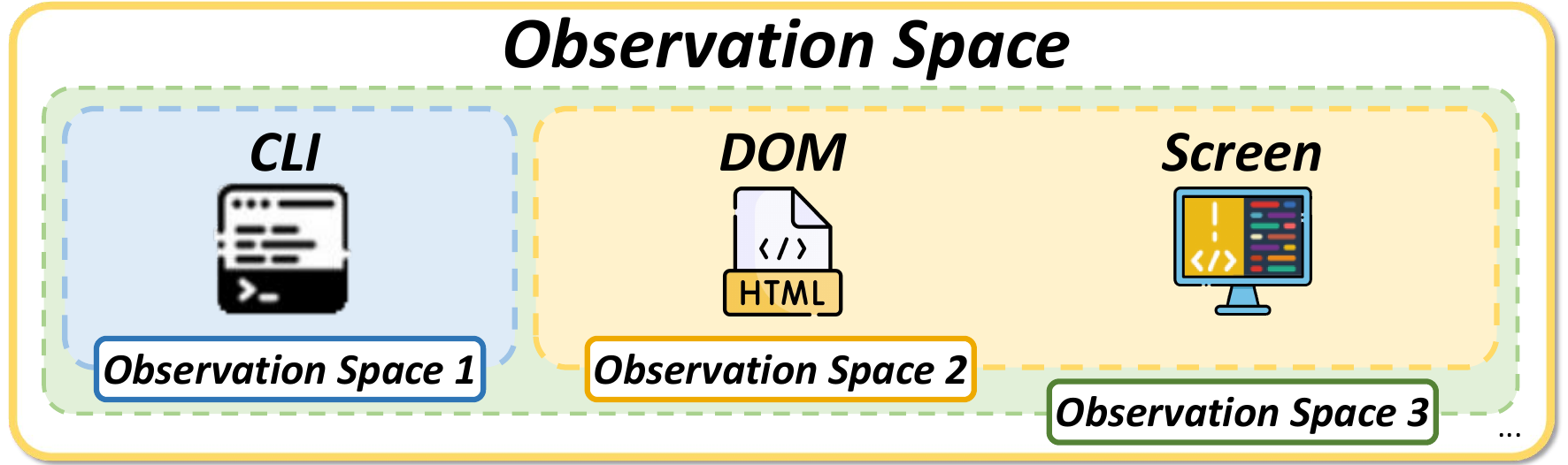}
\vspace{-5mm}
\centering\caption{Observation space is a single or combined representation of information, even the same environmental information will be presented to GVA differently through different observation spaces.}
\label{fig_5}
\vspace{-4mm}
\end{figure}

The observation space covers the range of information formats a GVA can perceive to effectively interpret its surroundings. There are currently three main forms: 1) \textbf{\textit{Command Line Interface (CLI)}}, where GVAs interpret system outputs and errors (Section~\ref{sec:2.3.1}); 2) \textbf{\textit{Document Object Model (DOM)}}, which allows GVAs to analyze web page structures (Section~\ref{sec:2.3.2}); 3) \textbf{\textit{Screen}}, where GVAs observe graphical states of interfaces (Section~\ref{sec:2.3.3}). We show some possible observation space combinations in Fig. ~\ref{fig_5}.

\subsubsection{\textbf{Command Line Interface}}
\label{sec:2.3.1}

CLI is a text-based UI that allows GVA to observe real-time feedback like compiler outputs and error messages. AutoGPT~\cite{Significant_Gravitas_AutoGPT} predefines code components, allowing the agent to observe execution results via the CLI. OSWORLD~\cite{OSWorld} and AGENTBENCH~\cite{DBLP:conf/iclr/0036YZXLL0DMYZ024} focus on OS tasks requiring CLI interaction. AgentStudio~\cite{zheng2024agentstudiotoolkitbuildinggeneral} supports creating reusable code scripts as tools and leveraging the CLI to use these tools. InterCode~\cite{DBLP:conf/nips/YangPNY23}, an interactive coding framework, allows agents to execute code in real-time via the CLI, providing feedback to enhance code quality.

The CLI provides a platform for GVAs to interact and learn, enabling the use of various tools or system interfaces that enhance their problem-solving capabilities. Although the CLI delivers valuable textual information, it cannot depict structural relationships between elements, thus limiting precise element grounding and hierarchical content understanding.

\subsubsection{\textbf{Document Object Model}}
\label{sec:2.3.2}

Systems like MiniWoB~\cite{ICLR17-Shi} and MiniWoB++~\cite{DBLP:conf/iclr/LiuGPSL18} use technologies like text feature maps map from query-DOM and DOMNET to interpret the DOM’s structure, while WebShop~\cite{NEURIPS2022_82ad13ec} and Mind2Web~\cite{NEURIPS2023_5950bf29} simplify data by filtering or selecting top-k relevant DOM elements. AppAgent~\cite{zhang2023appagentmultimodalagentssmartphone} combines XML with labeled screenshots for improved element recognition. CC-Net~\cite{DBLP:conf/icml/HumphreysRPTCMA22} uses both screenshots and DOM data to achieve human-level accuracy, but its performance drops significantly without DOM information, highlighting the DOM's importance for decision-making.

Though useful, access to structured data such as the DOM, is often restricted in sandbox environments or deliberately obscured by commercial websites to deter scraping, complicating data extraction and interpretation. Moreover, the length of HTML in real-world applications frequently exceeds the capabilities of standard language models, necessitating the use of advanced pruning algorithms to sift through and select essential data. These barriers impede the effectiveness and operational efficiency of GVAs in practical settings.

\subsubsection{\textbf{Screen}}
\label{sec:2.3.3}

In contrast to CLI and DOM, screen observation provides visual input, enabling GVAs to fully leverage the potential of Multimodal Large Language Models (MLLMs) for more precise perception. 
Agents like Auto-UI~\cite{zhang2024lookscreensmultimodalchainofaction}, SeeClick~\cite{cheng2024seeclickharnessingguigrounding}, WebVLN~\cite{chen2024webvln}, and Pix2Act~\cite{DBLP:conf/nips/ShawJCBPHKLT23} can perform tasks using only screenshots, without needing structured data. SeeAct~\cite{zheng2024gpt4visiongeneralistwebagent} noted that GPT-4V understands screens well but struggles with exact grounding. To improve accuracy, UFO~\cite{zhang2024ufouifocusedagentwindows} uses the Python package \textit{pywinauto} to label elements, while MM-Navigator~\cite{yan2023gpt} and Mobile-Agent~\cite{wang2024mobile} use OCR and icon detectors to enhance interaction.

Screen observation aligns with human intuition, facilitating natural interaction between GVAs and virtual environment. Acquiring screen observation is straightforward, requiring only a simple screenshot of the current screen state. The screen provides rich visual information, aiding in the accurate grounding and understanding of elements.

\begin{table*}[h!]
\caption{Comparison of various agents in terms of their components and reproducibility.}
\vspace{-2mm}
\centering
\Huge
\renewcommand{\arraystretch}{1.3}
\resizebox{\textwidth}{!}{%
\begin{tabular}{c|ccc|ccc|ccc|cccc|c}
\hline
\multirow{2}{*}{\textbf{Agents}} & \multicolumn{3}{c|}{\textbf{Environment}} & \multicolumn{3}{c|}{\textbf{Task}} & \multicolumn{3}{c|}{\textbf{Observation Space}} & \multicolumn{4}{c|}{\textbf{Action Space}} & \multirow{2}{*}{\textbf{Reproducibility}} \\ \cline{2-14}
 & Web & Application & OS & Command & Query & Dialogue & CLI & DOM & Screen & Keyboard & Mouse & Touchscreen & Others &  \\ \hline
MiniWoB~\cite{ICLR17-Shi} & \tick & \cross & \cross & \tick & \cross & \cross & \cross & \tick & \tick & \tick & \tick & \cross & \cross & \tick \\ 
MiniWoB++~\cite{DBLP:conf/iclr/LiuGPSL18} & \tick & \cross & \cross & \tick & \cross & \cross & \cross & \tick & \tick & \tick & \tick & \cross & \cross & \tick \\ 
CC-Net~\cite{DBLP:conf/icml/HumphreysRPTCMA22} & \tick & \cross & \cross & \tick & \cross & \cross & \cross & \tick & \tick & \tick & \tick & \cross & \cross & \cross \\ 
AppAgent~\cite{zhang2023appagentmultimodalagentssmartphone} & \cross & \tick & \cross & \tick & \cross & \cross & \cross & \tick & \tick & \cross & \cross & \tick & \cross & \tick \\ 
WebVoyager~\cite{he2024webvoyager} & \tick & \cross & \cross & \tick & \cross & \cross & \cross & \cross & \tick & \tick & \tick & \cross & \cross & \tick \\ 
Pix2Act~\cite{DBLP:conf/nips/ShawJCBPHKLT23} & \tick & \cross & \cross & \tick & \cross & \cross & \cross & \cross & \tick & \tick & \tick & \cross & \cross & \tick \\ 
Auto-UI~\cite{zhang2024lookscreensmultimodalchainofaction} & \cross & \cross & \tick & \tick & \tick & \cross & \cross & \cross & \tick & \cross & \cross & \tick & \cross & \tick \\ 
CogAgent~\cite{hong2023cogagentvisuallanguagemodel} & \cross & \cross & \tick & \tick & \tick & \cross & \cross & \cross & \tick & \tick & \tick & \tick & \cross & \tick \\ 
Mobile-Agent~\cite{wang2024mobile} & \cross & \tick & \cross & \tick & \cross & \cross & \cross & \cross & \tick & \cross & \cross & \tick & \cross & \tick \\ 
MM-Navigator~\cite{yan2023gpt} & \cross & \cross & \tick & \tick & \tick & \cross & \cross & \cross & \tick & \cross & \cross & \tick & \cross & \cross \\ 
AesopAgent~\cite{wang2024aesopagentagentdrivenevolutionarystorytovideo} & \cross & \tick & \cross & \tick & \cross & \cross & \cross & \cross & \tick & \tick & \cross & \cross & \cross & \cross \\ 
UFO~\cite{zhang2024ufouifocusedagentwindows} & \cross & \cross & \tick & \tick & \cross & \cross & \cross & \cross & \tick & \tick & \tick & \cross & \cross & \tick \\ 
SeeClick~\cite{cheng2024seeclickharnessingguigrounding} & \cross & \cross & \tick & \tick & \cross & \cross & \cross & \cross & \tick & \tick & \tick & \tick & \cross & \tick \\ 
AutoGPT~\cite{Significant_Gravitas_AutoGPT} & \cross & \cross & \tick & \tick & \cross & \cross & \tick & \cross & \cross & \tick & \cross & \cross & \cross & \tick \\ 
MMAC-Copilot~\cite{song2024mmaccopilotmultimodalagentcollaboration} & \cross & \cross & \tick & \tick & \tick & \cross & \tick & \cross & \tick & \tick & \tick & \cross & \cross & \cross \\ 
META-GUI~\cite{sun-etal-2022-meta} & \cross & \tick & \cross & \tick & \tick & \cross & \cross & \cross & \tick & \cross & \cross & \tick & \cross & \tick \\ \hline
\end{tabular}%
}
\label{table_1}
\vspace{-3mm}
\end{table*}

\subsection{\textbf{Action Space}}
\label{sec:2.4}

Action Space defines the range of interactive methods available to GVAs, allowing them to influence their environments. Reflecting human interactions with digital devices, the Action Space is categorized into four main modalities after a thorough literature review: 1) \textbf{\textit{Keyboard}} (Section~\ref{sec:2.4.1}), 2) \textbf{\textit{Mouse}} (Section~\ref{sec:2.4.2}), 3) \textbf{\textit{Touchscreen}} (Section~\ref{sec:2.4.3}), and 4) \textbf{\textit{Others}} (Section~\ref{sec:2.4.4}).

\subsubsection{\textbf{Keyboard}}
\label{sec:2.4.1}

The keyboard serves as a fundamental text input device for GVAs, enabling basic string input capabilities that facilitate interaction with digital environments. In early implementations like MiniWoB~\cite{ICLR17-Shi}, keyboard operations are limited to pressing one key at a time, resulting in low efficiency. Mind2Web~\cite{NEURIPS2023_5950bf29}, WebVoyager~\cite{he2024webvoyager}, and WebArena~\cite{DBLP:conf/iclr/ZhouX0ZLSCOBF0N24} improve efficiency by allowing multiple key presses per operation. WorkArena~\cite{workarena2024} expands operations to include string typing and distinguishing between key press and release actions.

The keyboard endows the GVA with basic text input capabilities, facilitating the interaction between the GVA and the environment. By inputting code into the environment, the GVA is capable of handling more complex tasks.

\subsubsection{\textbf{Mouse}}
\label{sec:2.4.2}

The mouse, as a core interactive device in graphical user interfaces, complements the keyboard by enabling the GVA to interact with the computer more intuitively. In MiniWoB~\cite{ICLR17-Shi}, five mouse operations are defined: no operation, click, drag, scroll up, and scroll down, with clickable coordinates set to [0, 160) × [0, 160). Pix2Act~\cite{DBLP:conf/nips/ShawJCBPHKLT23} refines this by discretizing coordinates to [0, 32) × [0, 32). WebVoyager~\cite{he2024webvoyager}, WebArena~\cite{DBLP:conf/iclr/ZhouX0ZLSCOBF0N24}, and Mind2Web~\cite{NEURIPS2023_5950bf29} enhance accuracy by limiting clickable positions to web elements. WorkArena~\cite{workarena2024} adds distinctions between mouse press and release actions.
The combined use of the keyboard and mouse provides the GVA with a powerful interaction method. This combination significantly enhances the GVA's task execution capabilities, enabling it to handle more complex tasks in a manner that closely resembles natural human interaction.

\subsubsection{\textbf{Touchscreen}}
\label{sec:2.4.3}

The introduction of touchscreen operations enables the GVA to interact with mobile devices, expanding its control beyond computers to smartphones, tablets, and smartwatches. In AppAgent~\cite{zhang2023appagentmultimodalagentssmartphone}, actions like tap, long press, and swipe are defined. AITW~\cite{NEURIPS2023_bbbb6308} and Auto-UI~\cite{zhang2024lookscreensmultimodalchainofaction} define a two-point gesture, merging drag and tap into one gesture. Mobile-Agent~\cite{wang2024mobile} specifies two tap operations: one for OCR-detected text and another for icon detection.

By defining a set of touchscreen operations, GVA can simulate human interactions with mobile devices. The diverse action space not only enhances the GVA's interaction capabilities but also provides a wider array of strategies and possibilities for executing complex tasks. As technology continues to advance, we can foresee that future action spaces will become richer and more intelligent, offering GVA more intuitive ways to interact.


\subsubsection{\textbf{Others}}
\label{sec:2.4.4}
Finally, we turn our attention to several specific scenarios.
For example: \textit{a) Gamepad}. A widely used input device for controlling robots in 2D space, offering intuitive and precise maneuverability. \textit{b) Gesture}. HyperPalm~\cite{DBLP:conf/smc/NazarovaBWFT22} introduces a gesture-based control method for 3D space, enabling natural interaction with quadruped robots through hand movements interpreted by advanced sensors. \textit{c) Voice Command Receiver}. This device uses speech recognition to control robots, ideal for hands-free operation. \textit{d) Brain-Computer Interface}. BCI allows direct brain-robot communication, interpreting neural signals for intuitive control, particularly beneficial for users with limited mobility or in complex decision-making scenarios.
They expand the operational domain of GVAs.

\section{Why do we need generalist virtual agents?}
\label{sec:3}
The impact of GVA extends far beyond its role as an intelligent assistant to humans; it is poised to offer immeasurable value in perspectives such as AI and machine learning, interaction, and practical applications. In this section, we will analyze the reasons why GVA is urgently needed.

\subsection{\textbf{Perspective of AI and Machine Learning}}
\label{sec:3.1}

The GVA demonstrates an adaptive capacity within complex and dynamic environments that surpasses the limitations of current large-scale models, which are typically confined to black-box functionalities accessible only through APIs. This adaptability is crucial for designing AI systems that perform optimally in real-world scenarios. Consequently, prioritizing the development of GVAs is essential. Beyond alleviating the complex learning burdens and inefficiencies currently borne by humans, GVAs introduce a novel approach in the realm of machine learning known as "grounded language learning." Unlike traditional models that merely align labels with virtual symbols, GVAs engage in learning through real-world observation and interaction~\cite{hong2023cogagentvisuallanguagemodel},~\cite{zhang2024lookscreensmultimodalchainofaction},~\cite{cheng2024seeclickharnessingguigrounding},~\cite{DBLP:conf/nips/ShawJCBPHKLT23},~\cite{zheng2024gpt4visiongeneralistwebagent},~\cite{wang2024mobile}. For example, GVAs understand the concept of "red" not just through static dictionary definitions but by observing a variety of red objects in situ, and they comprehend "heavy" by interacting with objects of varying weights. This shift from perception-based to comprehension-based learning signifies a fundamental transformation in the field of machine learning, enabling machines to not only process but genuinely understand and interact with their environments.


\subsection{\textbf{Perspective of Interaction}}
\label{sec:3.2}

Similar to the revolutionary introduction of sound in movies and the transition from command lines to graphical interfaces in computing, interaction modalities have evolved to align more closely with natural human behaviors such as listening and speaking. Despite these advancements, current manual operations require users to adapt to device-specific commands, often leading to inefficiency. This challenges us to think: why not enable machines to learn from the ``human manual"? Envisioning a new paradigm of human-machine interaction grounded in natural language dialogue, we propose employing a GVA. Equipped with deep comprehension abilities, GVA can redefine our interactions with technology. By providing feedback~\cite{DBLP:conf/acl/LinFKD22} tailored to individual preferences~\cite{DBLP:conf/nips/ChristianoLBMLA17} and providing personalized advice~\cite{Apple_Intelligence}, such agents could transform interactions to be more intuitive, making technology more accessible and aligning with natural human behaviors, thereby revolutionizing our engagement with devices.


\subsection{\textbf{Perspective of Agent Applications}}
\label{sec:3.3}

The efficacy of specialist tools in boosting computational efficiency is evident as they rapidly perform basic tasks like recognition, classification, and feedback. However, their efficiency requires human oversight and lacks integration across domains, showing that efficiency alone does not suffice for creative decision-making tasks. Therefore, there is a need for a GVA with advanced understanding and logical reasoning capabilities is essential to effectively address these challenges. Based on our analysis, we identify the following notable deficiencies in specialist tools: 1) Lack of coordinated combinatorial capability: independent tools cannot autonomously combine and arrange themselves to address more complex tasks. 2) Lack of real-time perception capabilities: tools are unable to perceive changes in the environment and adjust their outputs in response. 3) Unfriendly learning curve: most specialized tools require a significant investment in learning, presenting a high barrier to user proficiency. Given these limitations, there is an urgent need for a GVA endowed with planning capabilities~\cite{1316836},~\cite{kagaya2024rap},~\cite{baai2023plan4mc},~\cite{qiao2024autoact} that can self-update in real-time~\cite{NEURIPS2023_91edff07},~\cite{DBLP:conf/nips/ShinnCGNY23},~\cite{DBLP:conf/iclr/MaLWHBJZFA24} and offer a low threshold for user interaction~\cite{workarena2024},~\cite{wang2024aesopagentagentdrivenevolutionarystorytovideo}.

\section{How to implement generalist virtual agents?}
\label{sec:4}

In the previous section, we discussed the need for GVAs, highlighting their efficiency, their pioneering role in new machine-learning fields, and their potential to advance natural language interaction modes. Despite its significance, few papers detail its implementation. Thus, we outline GVA implementation from three aspects: 1) \textbf{\textit{Environment}}. A virtual space that GVA can \textbf{perceive and influence} when completing tasks (Section~\ref{sec:4.1}). 2) \textbf{\textit{Model}}. The brain of GVA, regarded as \textbf{a tunable function} that receives observations and returns actions (Section~\ref{sec:4.2}). 3) \textbf{\textit{Strategy}}. The methodologies for \textbf{training and orchestrating} the model to achieve optimal performance (Section~\ref{sec:4.3}). Our survey comprehensively considers the entire process of training a GVA from scratch.

\begin{figure}[t]
\includegraphics[width=\linewidth]{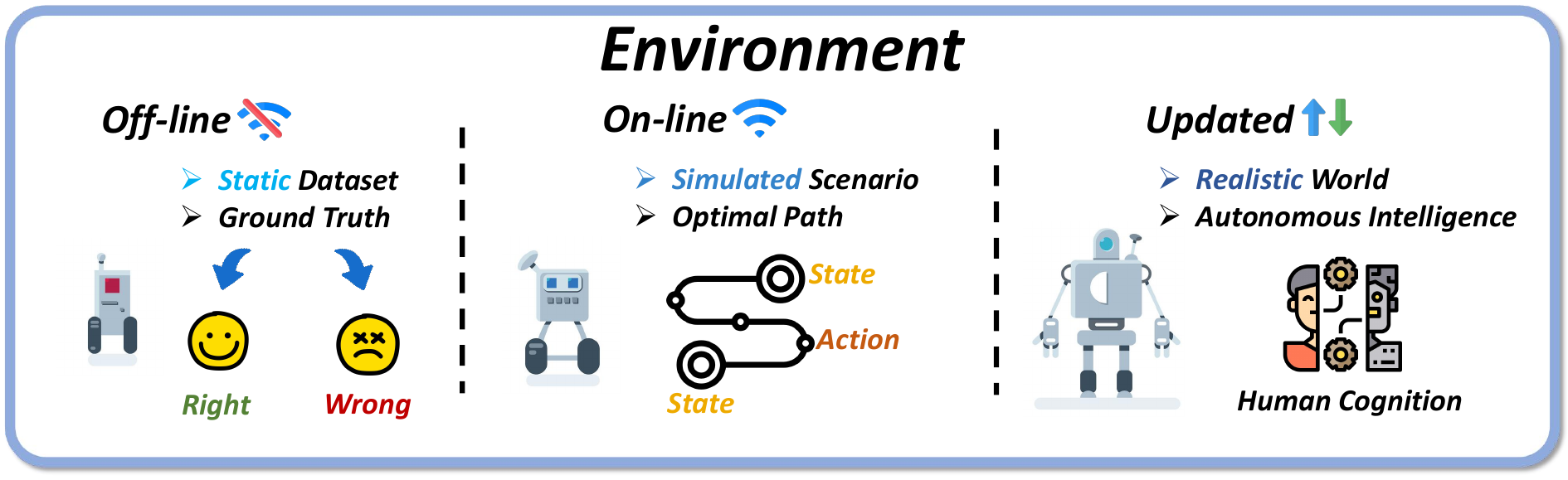}
\vspace{-5mm}
\centering\caption{We categorize environments based on their closeness to reality into three types: 1) Offline: a static dataset environment where learning is targeted at fitting the ground truth, 2) Online: a simulated website environment aimed at learning optimal pathways, and 3) Updated: a real-world environment where the learning objective is aligned with human cognition.}
\label{fig_6}
\vspace{-2mm}
\end{figure}

\subsection{\textbf{Environment}}
\label{sec:4.1}
The environment is a virtual space that GVA can perceive and influence when completing tasks. After reviewing numerous papers, we classify environments into three categories based on their authenticity, as shown in Fig.~\ref{fig_6}: 1) \textbf{\textit{Off-line}}, which has poor authenticity and typically contains static web pages or datasets (Section~\ref{sec:4.1.1}). 2) \textbf{\textit{On-line}}, which has better authenticity but still is a simulation of reality (Section~\ref{sec:4.1.2}). 3) \textbf{\textit{Updated}}, which is an authentic environment that includes unexpected events (Section~\ref{sec:4.1.3}). Next, we will introduce each type of environment separately.

\subsubsection{\textbf{Off-line}}
\label{sec:4.1.1}

The off-line environment features cached websites or static datasets, lacking authenticity and explorability. Constructing this environment simply involves storing all observations and actions in sequence as ground truth. This method efficiently gathers substantial data, enhancing the GVA's ability to follow instructions.

The environment in MiniWoB~\cite{ICLR17-Shi} and MiniWoB++~\cite{DBLP:conf/iclr/LiuGPSL18} uses cached websites. The environment in SeeClick~\cite{cheng2024seeclickharnessingguigrounding} collects GUI grounding data with instructions on various GUIs. The environment in AITW~\cite{NEURIPS2023_bbbb6308} collects task demonstration datasets on four Android systems and eight screen resolutions. Mind2Web~\cite{NEURIPS2023_5950bf29} collects snapshots from 137 websites and corresponds to over 2000 tasks. MoTIF~\cite{burns2022motifvln} is a static dataset with task feasibility annotations. The environment in META-GUI~\cite{sun-etal-2022-meta} consists of static observation-action trajectories.

GVA accesses only saved states in this environment, leading to task failure upon unexpected actions. Unlike reality, where multiple solutions exist for a task, this setup is restrictive. Thus, we will introduce an on-line environment that offers a more interactive and exploratory setting.

\subsubsection{\textbf{On-line}}
\label{sec:4.1.2}
The on-line environment offers a more realistic scenario compared to the off-line environment. It is dynamic and explorable, allowing GVAs to freely interact, and try various strategies. Typically based on simulations of real websites, the On-line environment provides GVAs with diverse action trajectories to complete tasks, making it ideal for developing their ability to generate varied solutions.

WebShop~\cite{NEURIPS2022_82ad13ec} is a simulation website for online shopping, containing 1.18 million real products crawled from Amazon. Agents can freely explore better task-solving strategies in this interactive environment. WebArena~\cite{DBLP:conf/iclr/ZhouX0ZLSCOBF0N24} has set up four simulation websites of different types, OneStopShop, CMS, Reddit and GitLab, covering four popular fields on the Internet: online shopping, content management, social forums, and collaborative development. VisualWebArena~\cite{koh2024visualwebarena} focuses on visual information in user instructions and has developed simulation implementations for websites such as OsClass, Reddit, and OneStopShop based on WebArena~\cite{DBLP:conf/iclr/ZhouX0ZLSCOBF0N24}. Agents can also explore various functions within websites and discover different task solutions.

Although the on-line environment is carefully crafted and freely explorable, it simplifies reality by excluding unexpected events which are common in real-life scenarios. To fully assess GVA performance in complex environments, we must consider more authentic settings.

\subsubsection{\textbf{Updated}}
\label{sec:4.1.3}

The updated environment is an authentic scenario comprising real websites and applications. It includes challenges like advertisements, pop-ups, captchas, and content updates over time, which test GVA's generalization abilities. Training in these real-life scenarios can significantly enhance GVA's capacity to handle unexpected events.

The environment in WorkArena~\cite{workarena2024} is based on the real website ServiceNow, and the environment in WebVoyager~\cite{he2024webvoyager} includes 15 real websites such as Google, Arxiv, and Apple. AppAgent~\cite{zhang2023appagentmultimodalagentssmartphone} and Mobile-Agent~\cite{wang2024mobile} use ADB tools to control Android devices, including tasks in common apps such as Gmail, YouTube, and Telegram. OSWORLD~\cite{OSWorld}, AgentStudio~\cite{zheng2024agentstudiotoolkitbuildinggeneral}, and UFO~\cite{zhang2024ufouifocusedagentwindows} all control computers, including complex tasks in real-world applications such as PowerPoint, Chrome, and VS Code.

These real-world environments pose challenges for GVA. Developing robust capabilities for GVAs in such settings is a crucial future research direction. This effort represents not only a technical breakthrough but also a vital step towards advancing AI to higher-level agents.

\subsection{\textbf{Model}}
\label{sec:4.2}
As an autonomous agent, GVA requires a versatile ``core brain'' for understanding instructions and making decisions. Based on the characteristics of these brain models, they can be categorized as follows: 1) \textbf{\textit{Retriever-based Agent}}: An agent that utilizes a core retrieval system to process natural language and images. 2) \textbf{\textit{LLM-based Agent}}: An agent powered by language models, specializing in processing natural language instructions. 3) \textbf{\textit{MLLM-based Agent}}: An agent that employs multimodal language models, handling both textual and visual instructions. 4) \textbf{\textit{VLA-based Agent}}: An agent based on Vision-Language-Action (VLA) models, integrating visual and linguistic inputs with action outputs. Each model type exhibits distinct characteristics and advantages, tailored to meet diverse interaction and processing requirements. The representative work of each model is shown in Table~\ref{table_2}.

\subsubsection{\textbf{Retriever-based Agent}}
\label{sec:4.2.1}
This agent relies on an advanced retrieval system to efficiently extract and process user-requested information from large data repositories. It combines search engine technologies with basic NLP capabilities to understand user queries and deliver relevant responses, suitable for routine queries and simple tasks.

R-MAGIC~\cite{1316836} features an agent-based multimedia retrieval architecture that integrates multiple repositories into a unified system. Here, agents collaboratively develop retrieval strategies tailored to various datasets and specific needs. Additionally, Mobile Agents~\cite{980543} utilize mobile agents to extract data from electronic calendars, optimizing scheduling for multi-party events and enhancing efficiency, privacy, and system load management. Furthermore, SAIRE~\cite{odubiyi1997saire}, a scalable agent-based information retrieval engine, improves upon traditional search engines by using intelligent agents and NLP techniques for enhanced retrieval capabilities, supporting conceptual searches and accommodating user preferences. Lastly, ACQUIRE~\cite{das2002acquire} simulates a unified interface for homogeneous data sources, processing queries by translating them into multiple sub-queries, with mobile agents retrieving and returning data from designated sources.

The retriever-based Agent model excels in rapid response and real-time query management, ideal for applications like live customer support. However, its dependence on content retrieval can limit inferential and personalization abilities, sometimes leading to inadequate responses. With the rise of LLMs, retrievers have transitioned to supportive roles, as demonstrated by the Thought-Retriever~\cite{feng2024thought}, heralding a new era dominated by LLM-based Agents.

\begin{table}[]
\begin{center}
    \captionsetup{font={small, stretch=1.25}, labelfont={bf}}
    \caption{GVA model categories and their respective base models.}  
    \vspace{-4mm}
    \renewcommand{\arraystretch}{1.2}
    \resizebox{0.5\textwidth}{!}{

    \begin{tabular}{c|l|c}
    \toprule[1pt]

    \multicolumn{1}{c|}{Type}            & Work           & Model             \\
    \hline \hline
    \multirow{4}{*}{Retriever-based Agent} & R-MAGIC~\cite{1316836}        & /                 \\
                                        & Mobile Agents~\cite{980543}  & /                 \\
                                        & SAIRE~\cite{odubiyi1997saire}          & /                 \\
                                        & ACQUIRE~\cite{das2002acquire}        & /                 \\
    \hline
    \multirow{10}{*}{LLM-based Agent}      & Agent-Pro~\cite{zhang2024agent}      & GPT-4             \\
                                        & LOGIC-LM~\cite{DBLP:conf/emnlp/PanAWW23}       & GPT-4             \\
                                        & HuggingGPT~\cite{DBLP:conf/nips/0001ST00Z23}     & GPT-4             \\
                                        & MM-REACT~\cite{yang2023mm}       & ChatGPT           \\
                                        & CoELA~\cite{DBLP:conf/iclr/ZhangDSZDTSG24}          & GPT-4             \\
                                        & RoboGPT~\cite{chen2024robogptintelligentagentmaking}        & ChatGPT           \\
                                        & AgentCoder~\cite{huang2024agentcodermultiagentbasedcodegeneration}     & GPT-4             \\
                                        & Pyramid Coder~\cite{shen2024pyramid}  & GPT-3.5; CodeLlama \\
                                        & CodeVQA~\cite{DBLP:conf/acl/SubramanianNKYN23}        & BLIP              \\
                                        & ViperGPT~\cite{DBLP:conf/iccv/SurisMV23}       & GPT-3; CLIP        \\
    \hline
    \multirow{10}{*}{MLLM-based Agent}     & MM-Navigator~\cite{yan2023gpt}   & GPT-4             \\
                                        & Mobile-Agent~\cite{wang2024mobile}   & GPT-4             \\
                                        & AER~\cite{pan2024autonomous}            & GPT-4             \\
                                        & CLOVA~\cite{gao2024clova}          & ChatGPT           \\
                                        & Agent Smith~\cite{gu2024agent}    & ChatGPT           \\
                                        & WebVoyager~\cite{he2024webvoyager}     & GPT-4             \\
                                        & Auto-UI~\cite{zhang2024lookscreensmultimodalchainofaction}        & GPT-3.5; CodeLlama \\
                                        & VisualWebArena~\cite{koh2024visualwebarena} & BLIP              \\
                                        & CogAgent~\cite{hong2023cogagentvisuallanguagemodel}       & GPT-3; CLIP        \\
    \hline
    \multirow{6}{*}{VLA-based Agent}       & Rt-2~\cite{brohan2023rt}            & PaLI-X; PaLM-E     \\
                                        & Lm-nav~\cite{shah2023lm}         & CLIP              \\
                                        & Octo~\cite{team2024octo}           & PaLI-X; PaLM-E     \\
                                        & OpenVLA~\cite{kim2024openvla}        & LLaVA; IDEFICS-1   \\
                                        & 3d-vla~\cite{zhen20243d}         & 3D-LLM            \\
                                        & Actra~\cite{ma2024actra}          & ChatGPT; RT-1     \\
    \hline
    \end{tabular}

    }
    \label{table_2}
\end{center} 
\vspace{-3mm}
\end{table}

\subsubsection{\textbf{LLM-based Agent}}
\label{sec:4.2.2}
LLMs have shown capabilities nearing human intelligence, prompting extensive research into autonomous agents driven by these models. LLMs are favored for their exceptional ability to understand and generalize from textual inputs. Their core advantage lies in adeptly handling natural language, grasping contextual nuances and semantics, and producing coherent, logically consistent responses.

Utilizing GPT through APIs, various projects tailor distinct observation spaces to their specific tasks. For example, projects like Agent-Pro~\cite{zhang2024agent}, LOGIC-LM~\cite{DBLP:conf/emnlp/PanAWW23}, HuggingGPT~\cite{DBLP:conf/nips/0001ST00Z23}, and MM-REACT~\cite{yang2023mm} provide LLMs with natural language descriptions. Conversely, initiatives such as Mind2Web~\cite{NEURIPS2023_5950bf29} and AllTogether~\cite{liu2023alltogetherinvestigatingefficacyspliced} utilize HTML to create an understanding of web layouts and functionalities, parsing tags and attributes to minimize misoperations due to linguistic ambiguities. Open-source models like Llama are used in systems like CoELA~\cite{DBLP:conf/iclr/ZhangDSZDTSG24}, which provide textual tasks and image captions to address LLMs’ limitations with visual inputs. RoboGPT~\cite{chen2024robogptintelligentagentmaking} applies Llama in decomposing tasks for completing complex embodied tasks.

To meet the demands of complex operations and advanced automation scenarios, projects such as AgentCoder~\cite{huang2024agentcodermultiagentbasedcodegeneration}, Pyramid Coder~\cite{shen2024pyramid}, CodeVQA~\cite{DBLP:conf/acl/SubramanianNKYN23}, and ViperGPT~\cite{DBLP:conf/iccv/SurisMV23} adopt code generation models including CodeX, CodeLlama, and AlphaCode. These models are employed to perform more complex modular tasks, maximizing the scalability of code. Despite a steep learning curve that may reduce user-friendliness, the intrinsic interpretability and adeptness in managing intricate tasks render code pre-training models indispensable in advanced automation scenarios.

LLM-based agents excel at understanding and generating complex linguistic structures, making them ideal for various conversational contexts. Besides mastering natural language, these agents effectively manage structured text, facilitating dynamic content generation on digital platforms. However, the text-centric nature of LLMs, constrained by context window length and dependence on structured text, does not fully match the intuitive interaction typical of human behavior. This limitation highlights the need for vision-based agent models, which could handle a wider range of tasks and offer more natural interaction logic.

\subsubsection{\textbf{MLLM-based Agent}}
\label{sec:4.2.3}
The MLLM-based agent enhances LLM by processing multiple data modalities, capable of handling text and relevant information from other modalities. Its multimodal capabilities allow for effective user interactions in diverse contexts, processing inputs and producing cross-modal outputs to meet complex user demands.

Current research leveraging GPT-4V includes projects like MM-Navigator~\cite{yan2023gpt}, Mobile-Agent~\cite{wang2024mobile}, and AER~\cite{pan2024autonomous}. CLOVA~\cite{gao2024clova} operates on the BLIP model, while Agent Smith~\cite{gu2024agent} utilize the LLaVA framework. These initiatives significantly expand the observation space for agents by integrating textual instructions with screen captures or video clips as inputs. Additionally, projects such as WebVoyager~\cite{he2024webvoyager}, Auto-UI~\cite{zhang2024lookscreensmultimodalchainofaction}, VisualWebArena~\cite{koh2024visualwebarena}, and CogAgent~\cite{hong2023cogagentvisuallanguagemodel} incorporate the HTML as an additional input modality. This provides agents with a comprehensive view, highlighting interactive elements that are obscured in purely textual or image-based inputs. This enhancement makes GVAs more efficient and accurate in processing web content.

The integration of multimodality in agent interactions marks a pivotal shift towards a more human-like approach in digital environments. This advancement allows agents to interpret complex content in images, discern voice tones and emotions, and analyze dynamic video scenes, leading to more precise and personalized responses. Such capabilities not only enhance comprehension but also significantly expand their application range, particularly in accessibility technologies. Multimodality facilitates complex input methods like voice or touch, aligning with human intuitive interaction logic. Consequently, MLLM-based agents have become leaders in agent evolution.

\subsubsection{\textbf{VLA-based Agent}}
\label{sec:4.2.4}
The VLA-based agent combines visual processing, language comprehension, and action decision-making. It analyzes visual data, communicates via natural language with humans or systems, and executes actions based on the synthesized information. Its applications include service robotics, autonomous vehicles, and interactive systems, enhancing their ability to intelligently perceive and respond to their environment. Moreover, it equips these systems to handle increasingly complex and dynamic tasks, adapting to environmental changes with high efficiency.

These agents demonstrate capabilities for perceiving and interacting with real-world environments. Projects such as Rt-2~\cite{brohan2023rt}, Lm-nav~\cite{shah2023lm}, Octo~\cite{team2024octo}, and OpenVLA~\cite{kim2024openvla} showcase their ability to modify physical environments through visual perception, both in simulated and actual environments. They are particularly effective in spatial localization and physical sensing, facilitating accurate physical predictions from images and videos. Furthermore, research initiatives like 3d-vla~\cite{zhen20243d}, PaLM-E~\cite{driess2023palm}, and Actra~\cite{ma2024actra} are advancing models and frameworks for 3D environments, marking a transition towards embodied agents, representing a promising and evolving paradigm with significant potential for future applications.

The VLA-based agent boosts adaptability and efficiency in complex environments through integrated decision-making across visual, linguistic, and action domains. Directly interacting with environments to gather data and perform tasks, these agents exemplify perfect integration of perception and action. They support the development of embodied agents in both real and simulated settings, enabling them to process data, respond to commands, and autonomously refine actions based on physical interactions, thus enhancing adaptiveness and autonomy in intricate environments.

\subsection{\textbf{Strategy}}
\label{sec:4.3}
\begin{figure}[t]
\includegraphics[width=\linewidth]{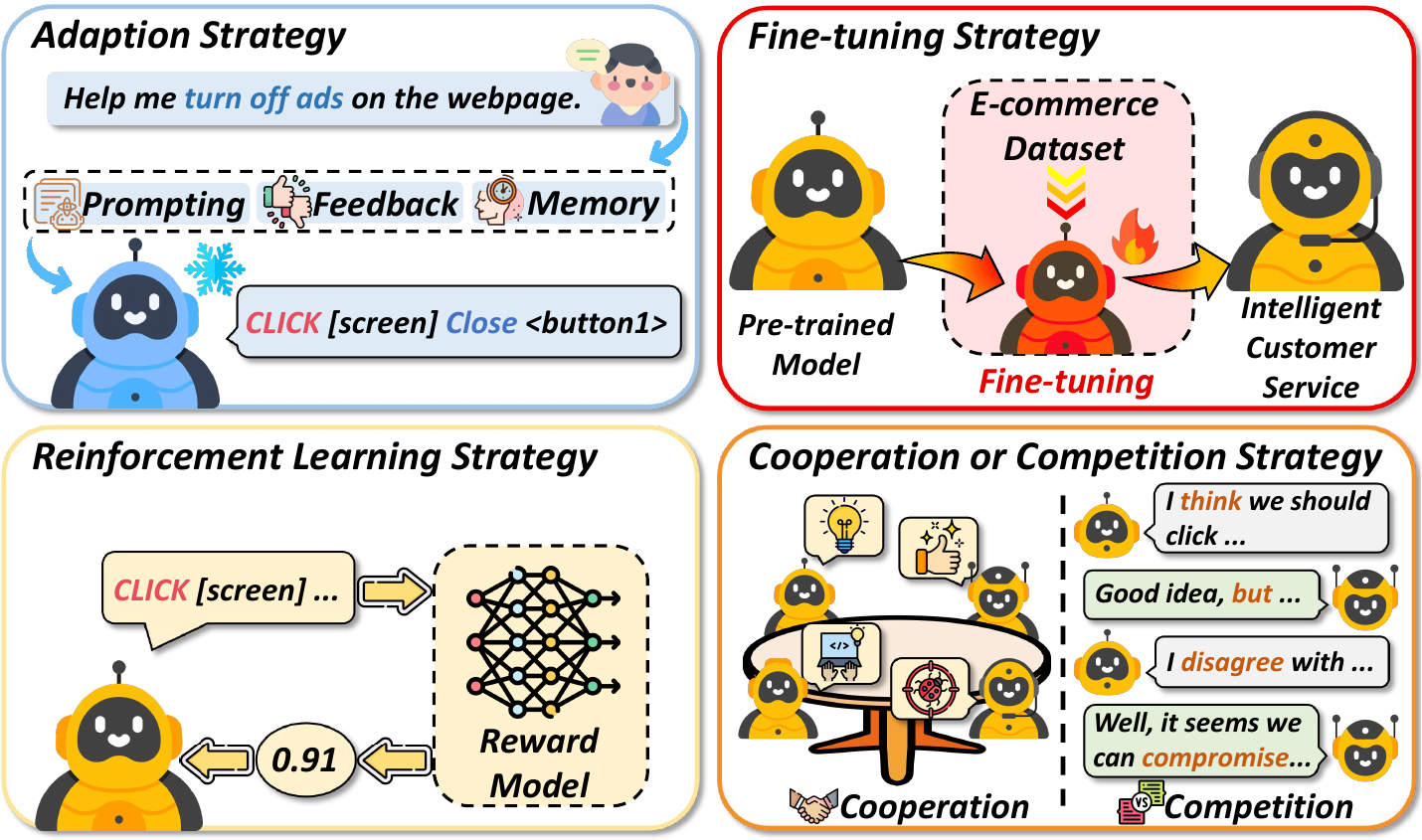}
\vspace{-5mm}
\centering\caption{Four GVA implement strategies with their respective characteristics.}
\label{fig_7}
\vspace{-2mm}
\end{figure}

In constructing a GVA, selecting the right strategy is essential for aligning with the tasks and interacting between the agent and users. Based on existing research, the strategies can be grouped into four main types (Fig.~\ref{fig_7}): 1) An \textbf{\textit{adaption strategy}} that operates without the whole model training, 2) A \textbf{\textit{fine-tuning strategy}} that requires supervisory signals for optimization, 3) A \textbf{\textit{reinforcement learning strategy}} that functions without supervisory signals, and 4) A \textbf{\textit{cooperation or competition strategy}} that engages in multimodal collaboration or human-machine interactions. We will introduce them respectively.

\subsubsection{\textbf{Adaption strategy}}
\label{sec:4.3.1}
Large-scale models exhibit generalization and zero-shot abilities, performing well with minimal contextual demonstrations or a simple task description. Termed an ``adaption strategy'', it bypasses model training or internal parameter modifications, instead adjusting prompts or training a lightweight layer structure to suit various tasks, offering significant flexibility.

\textbf{Prompting} strategy involves supplying a detailed task description or contextual examples to an agent model, thereby enabling it to initiate appropriate actions. Notable implementations include  MM-REACT~\cite{yang2023mm}, which uses prompting to engage ChatGPT in multimodal reasoning and action; Plan-and-Solve Prompting~\cite{wang2023plan} enhancing zero-shot task performance via a thought-chain prompt; and ExpertPrompting~\cite{xu2023expertprompting}, exploiting the potential of LLMs to serve as expert responders. These strategies involve merely providing prompts to pre-existing large models, eliminating the need for further training and offering a straightforward method for deploying agents.

\textbf{Feedback} strategy integrates outcomes from itself or other agents to refine initial responses, fostering model self-update or cooperation and enhancing understanding of tasks. Strategies like Self-Refine~\cite{madaan2024self} leverage a self-iterative mechanism involving feedback and refinement steps to deliver strong performance in natural language tasks. PanGu-Coder2~\cite{shen2023pangu} employs a feedback ranking system for better code generation. A prominent embodiment is De-fine~\cite{gao2023fine}, which incorporates four types of feedback—visual, textual, compile, and human—to elevate agent capabilities in multimodal tasks.

\textbf{Memory} strategy in intelligent agents is managed through a `store-process-action'~\cite{zhang2024survey} mechanism that selectively filters and enhances information. This approach allows agents to retain past experiences and data, facilitating better decision-making and adaptability in subsequent tasks. Systems like RAP~\cite{kagaya2024rap} use a retrieval-augmented strategy for contextual memory in MLLMs, while ChatDB~\cite{hu2023chatdb} handles database tasks with symbolic memory. MemoryBank~\cite{zhong2024memorybank}, inspired by the Ebbinghaus Forgetting Curve theory, enables AI to manage memories based on their relevance and elapsed time, simulating human-like memory processes.

Adaption strategies are valued for their simplicity, as they do not require retraining and maintain consistent model parameters as tasks vary, with only adjusting prompting templates or processes. However, they necessitate clear task descriptions, making the design process laborious, and rely on large, generic models that lack specialized knowledge for specific tasks and struggle with multi-application integration. The next section will discuss strategies trained on specific datasets.

\subsubsection{\textbf{Fine-tuning strategy}}
\label{sec:4.3.2}
It starts with a pre-trained model, followed by additional training on a specific smaller dataset to adjust the model’s parameters for better task-specific adaptation. This approach allows agents to use the broad capabilities of the pre-trained model while making targeted adjustments to improve performance on specific tasks.

Fine-tuning necessitates high-quality data, with current methodologies primarily focused on improving the quality of learning samples. The COEVOL~\cite{li2024coevol} devises a debate-advise-edit-judge cycle to iteratively refine responses tailored for fine-tuning. Re-ReST~\cite{dou2024reflection} integrates a `reflector' during its self-training phase to improve sample quality by incorporating feedback from external environments. ACT~\cite{chen2024learning} utilizes contrastive sample creation to sharpen the model’s dialogue strategies, enabling optimal response selection across diverse contexts. Additionally, efforts involve using GPT-4 generated data to improve interaction behaviors and minimize errors~\cite{zhou2024enhancing}, aiming at minimizing format errors and hallucinatory outputs. Furthermore, Agent-FLAN~\cite{chen2024agent} improves training datasets through functional decoupling and negative sample construction.

Fine-tuning enhances pre-trained models' performance on specialized tasks, offering rapid adaptability and resource efficiency. These strategies outperform prompting methods and are more efficient than unsupervised reinforcement learning but require high-quality annotated data and risk overfitting. The success of fine-tuning is critically dependent on the careful selection of both the pre-trained models and representative training datasets. Therefore, a training methodology that does not rely on annotated data may be more suitable for agents.

\subsubsection{\textbf{Reinforcement learning strategy}}
\label{sec:4.3.3}
In GVA training, personalization often makes initial labeling impractical due to two primary factors: 1) Non-uniqueness, where multiple action sequences can accomplish the task, and 2) Ambiguity, where the objectives are not clearly defined at first. In such scenarios, employing a reinforcement learning strategy with a well-designed reward function facilitates the model’s exploration of the environment and identification of optimal action sequences. This method allows the GVA to adapt its behavior based on the effectiveness of different approaches, continuously refining its actions to better meet user needs.

For instance, RL4VLM~\cite{zhai2024fine} employs Vision-Language Models (VLMs) to generate actionable steps, engaging with environment to secure task-oriented rewards. These rewards subsequently inform the training of the entire VLM system. Simultaneously, GLAINTEL~\cite{fereidouni2024search} explores enhancing smaller language models' performance in web environments through reinforcement learning, without human demonstrations.

Reinforcement learning is widely used in gaming, notably by AlphaGo in Go competitions and in Minecraft via platforms like Juewu-mc~\cite{lin2021juewu}, Minerl 2020~\cite{guss2021towards}, and Plan4mc~\cite{baai2023plan4mc}. It's also used in Pokémon games through the PokeLLMon~\cite{hu2024pok} agent. Google has also developed a platform that trains a game agent across its extensive gaming library~\cite{abi2024scaling}. Furthermore, studies presented in papers such as LLM-X~\cite{lu2024mental}, and LLMSat~\cite{maranto2024llmsat} investigate the application of reinforcement learning algorithms to optimize communication and collaboration strategies among language agents within role-playing elements in agent systems.

Reinforcement learning strategies enable agents to optimize behavior through trial and error, ideal in environments where goals are unclear. By autonomously exploring, agents independently identify effective strategies without needing detailed human instructions. This reduces reliance on expert knowledge and increases the system's flexibility and scalability.

\subsubsection{\textbf{Cooperation or competition strategy}} 
\label{sec:4.3.4}
These strategies in multi-agent systems enable agents to interact and communicate in a shared environment aimed at common goals. Unlike multimodal collaboration, this approach allows each agent to access all necessary information and complete sub-tasks strategically, supporting the collective achievement of main objectives. These interactions may be cooperative or competitive, details of which will be discussed next.

CMAT~\cite{liang2024cmat}  is a multi-agent framework that enhances language agents by adapting based on environmental feedback, improving situational awareness and memory. AUTOACT~\cite{qiao2024autoact} features three agents—Plan, Tool, Reflect—and implements Group Planning with integrated strategies. ProAgent~\cite{zhang2024proagent} adapts to teammates' intentions and actions using various modules to predict behaviors accurately. As ChatLLM~\cite{hao2023chatllm} indicates, ``more brains, more intelligence''.

While MAD~\cite{liang2023encouraging}, Multiagent Debate~\cite{du2023improving}, ChatEval~\cite{chan2023chateval}, LOGICOM~\cite{payandeh2023susceptible} focus on developing intelligent agents through debate and competition. This interaction allows agents to challenge each other's ideas, and engaging in debates exposes agents to diverse perspectives and strategies, leading to innovative solutions and fewer errors due to individual biases, ultimately fostering a more robust evolution of capabilities than single-agent systems.

Multi-agent learning addresses information asymmetry by allowing each agent to independently make decisions toward a shared goal. This approach enables agents to specialize and excel in distinct domains, managing subtasks aligned with the collective objective. Mirroring the distribution of responsibilities in human societies, where collaboration achieves otherwise unattainable goals, this model highlights a promising direction for future GVA research focused on collaborative intelligence and strategic task distribution.

\begin{figure*}[t]
\includegraphics[width=1\textwidth]{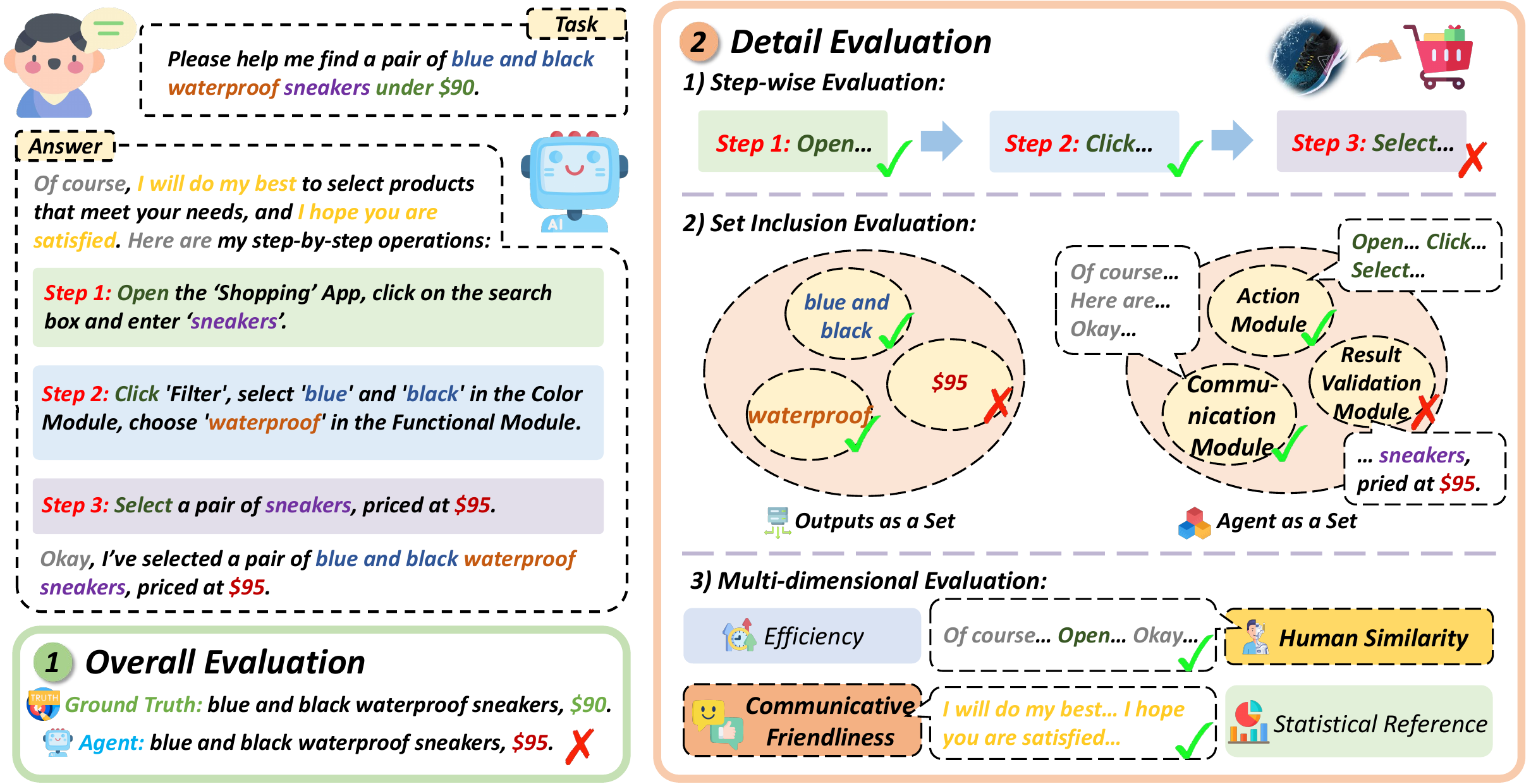}
\vspace{-5mm}
\centering\caption{We illustrate the distinctions between two evaluation approaches used in quantitative tasks: 1) overall evaluation: a coarse-grained binary classification that compares the output to the ground truth, and 2) detail evaluation: a fine-grained, multi-dimensional evaluation of processes and models. Additionally, this example demonstrates three different matrices used for detailed evaluation.}
\label{fig_8}
\vspace{-2mm}
\end{figure*}

\section{How to evaluate Generalist Virtual Agent?}
\label{sec:5}

The development of GVA expands their ability to handle numerous tasks. The methods for evaluating GVA must continuously evolve to comprehensively assess their ability of reasoning, planning, tool-using, and so on. Existing research offers various evaluation strategies based on different applications. However, these studies often focus on specific tasks or steps and lack standardized guidelines for evaluating agents' systematic and comprehensive performance.

\begin{figure*}[t]
\includegraphics[width=1\textwidth]{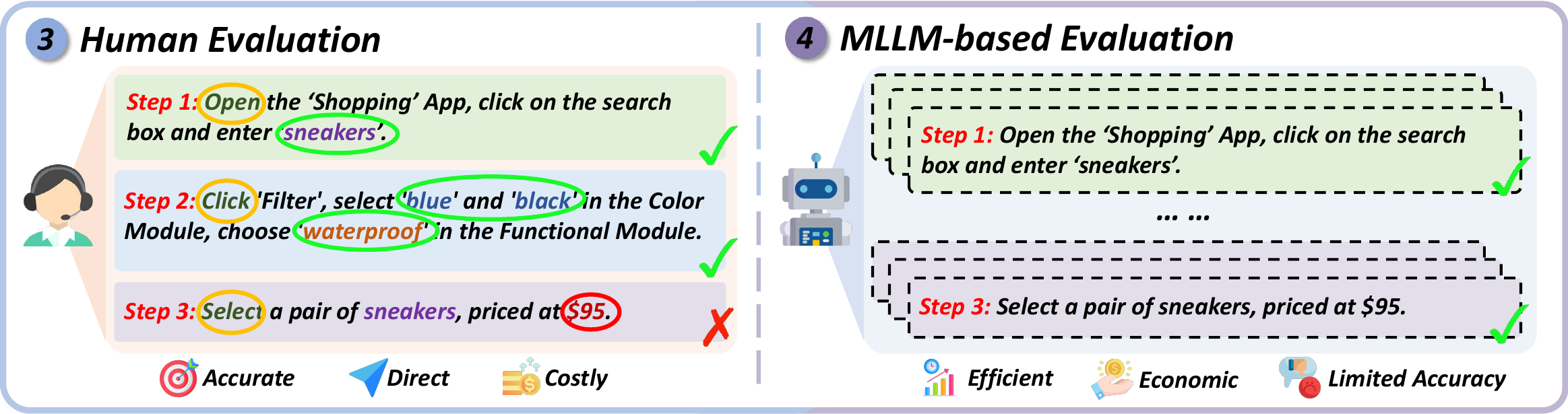}
\vspace{-5mm}
\centering\caption{In qualitative evaluation, human evaluation is often synonymous with high quality and irreproducibility. Correspondingly, model evaluations are characterized by their consistency and limited interpretability.}
\label{fig_9}
\vspace{-3mm}
\end{figure*}

In this section, we summarize the general evaluation framework for GVA as follows: A) \textit{\textbf{Overall Evaluation}}: A coarse-grained quantitative evaluation that compares the results with the ground truth (Section~\ref{sec:5.1}); B) \textit{\textbf{Detail Evaluation}}: A fine-grained, multi-dimensional quantitative evaluation of processes and modules (Section~\ref{sec:5.2}); C) \textit{\textbf{Human Evaluation}}: A direct qualitative evaluation based on high-quality human supervision (Section~\ref{sec:5.3}); D) \textit{\textbf{MLLM-based Evaluation}}: An efficient qualitative evaluation using intelligent multimodal large language models (Section~\ref{sec:5.4}). We use online shopping scenarios and different evaluation methods to highlight their characteristics in Fig.~\ref{fig_8} \&~\ref{fig_9}.

\subsection{\textbf{Overall Evaluation}}
\label{sec:5.1}

Overall evaluation is a \textbf{coarse-grained binary classification} that compares the output with the ground truth, suitable for general scenarios like task completion or benchmark assessments. It is straightforward and intuitive, directly informing users whether the agent has completed the task. Its simplicity also ensures effectiveness across various contexts without being influenced by specific datasets or model variations.

Specifically, many agents are evaluated based on the $Success\,Rate$ ($SR$) of task completion as a metric of whether the agent has completed the task, formulated as:
$$
SR=\frac{1}{N}\sum^N_{i=1}\mathbb{I}[a_i=\hat a_i]
$$
where $N$ is the number of questions, $a_i $ represents the ground truth of the $i$-th question, and $\hat a_i$ represents the output given by the agent. The indicator function $\mathbb{I}[·]$ returns 1 if the elements inside are equal, and 0 otherwise. We list some works that use overall evaluation in Table~\ref{tab:Classification_of_metrics}.

Overall evaluation often yields binary outcomes, $\{0,1\}$, but these results have limitations. A score of $1$ doesn't always signify efficiency; the agent might complete the task but at a high cost in time and resources. Conversely, a score of $0$ doesn't indicate a complete failure, as it can identify areas for improvement. Thus, relying solely on this method is neither reasonable nor prudent. A more detailed and comprehensive evaluation method, known as detail evaluation, is required.

\subsection{\textbf{Detail Evaluation}}
\label{sec:5.2}

Detail evaluation refers to a \textbf{fine-grained, multi-dimensional evaluation} of an agent's task completion process and specific modules. This approach is suitable for scenarios where the various abilities of the agent and task completion are evaluated. Furthermore, a fine-grained evaluation can concentrate on specific actions or sequences, enhancing interpretability and ensuring a more scientifically rigorous and justified evaluation process. After summarizing and organizing, we categorize the existing detail evaluation methods as follows:

\subsubsection{\textbf{Step-wise Evaluation}}
\label{sec:5.2.1}
It refers to the evaluation of the sequential action steps taken by the agent to complete a task. Suppose the number of correct steps executed by the agent is $Num_{mat}$, and the total number of steps is $Num_{sum}$. Step-wise score $StepScore_{div}$ can be defined as:
$$
StepScore_{div}=\frac{Num_{mat}}{Num_{sum}}
$$
In this type of evaluation method, different works define $Num_{sum}$ differently: Some works define $Num_{sum}$ as the total number of ground truth steps, reflecting the extent to which the agent's steps match human preferences~\cite{ICLR17-Shi},~\cite{NEURIPS2023_5950bf29},~\cite{li-etal-2020-mapping}. Other works define $Num_{sum}$ as the total number of steps taken by the agent~\cite{sun-etal-2022-meta},~\cite{NEURIPS2023_bbbb6308},~\cite{wang2024mobile}. In particular, there is also a study~\cite{chen-etal-2021-websrc} that defines $Num_{sum}$ as the union of the agent total steps and the actual steps. 

Due to the sequential nature of the steps, there is also an step-wise evaluation method based on a sequence of actions. WebVLN\cite{chen2024webvln} transforms the step-wise completion of tasks by the agent into a ``vision-language" navigation task, similar to Room2Room\cite{CSCW16-Pejsa}, which calculates the shortest path distance between the final located webpage and the target webpage. In OmniACT\cite{kapoor2024omniactdatasetbenchmarkenabling}, the sequence score ($SeqScore$) is defined as:
$$
SeqScore_i=\begin{cases}     \beta_1+\beta_2*(s-1), & \text{if all actions match} \\    0, & \text{otherwise}\end{cases}
$$
where $s$ is the length of the action sequence, $\beta_1$ and $\beta_2$ are hyperparameters, set to $0.1$ and $1$, respectively.

Step-wise evaluation offers a practical method for agents sequentially completing tasks. However, step-wise evaluation methods have limitations. For example, there isn't a single efficient path to complete a task. We also need to consider shortcut key operations on different platforms, as these operations can affect the quality of step-wise evaluation.

\subsubsection{\textbf{Set Inclusion Evaluation}}
\label{sec:5.2.2}
It refers to the evaluation of modules or output results of an agent as a set. This strategy aims to assess task completion by set inclusion —- the more correct components included, the more successful the task completion. Specifically: a) \textbf{Agent as a Set}: The agent is viewed as a set of modules. The agent completes tasks by combining modules, and each module will be assessed. b) \textbf{Outputs as a Set}: The agent's output results are viewed as a set of attributes. For tasks such as web content Q\&A, the attributes are compared to those in the ground truth.

In existing work, RUSS\cite{xu-etal-2021-grounding} evaluates the semantic parser and element localization module of the agent using $Exact\,Match\,Accuracy$ and $Grounding\,Accuracy$, respectively. META-GUI\cite{sun-etal-2022-meta} introduces the $BLEU\,Score$ to evaluate the Response Module of the agent. VisualWebArena\cite{koh2024visualwebarena} uses two metrics, $must\_include$, and $must\_exclude$, to represent the results of web content Q\&A, refining the evaluation by including certain elements and excluding others. WebQA\cite{Chang_2022_CVPR} introduces $BARTScore$ to consider semantic consistency and evaluate the language fluency of the answers.

Additionally, in WebShop\cite{NEURIPS2022_82ad13ec}, Yao et al. designed a reward function to evaluate the degree of conformity of the products purchased by the agent. Specifically, for a certain product $x$ that the user wants to purchase, it has a set of non-empty attributes $U_ {att} $, a set of specific option descriptions $U_ {opt} $, and an expected highest price $u_ {price} $. The item purchased by the agent is $y $, which has a corresponding set of non-empty attributes $Y_ {att} $, a specific option description $Y_ {opt} $, and a specific price $y_ {price} $. Therefore, the reward function is characterized as:
$$
r=r_{type}\cdot\frac{|U_{att}\cap Y_{att}|+|U_{opt}\cap Y_{opt}|+1[y_{price}\leq u_{price}]}{|U_{att}|+|U_{opt}|+1}
$$
where $r_ {type} $ is a type reward function value obtained by heuristic methods based on text matching.

Set inclusion evaluation is complex and has different standards. For instance, in web element grounding, some tasks assess whether the agent's predicted point lies within the ground truth bounding box\cite{NIPS14-Sutskever}, while others calculate the IoU (Intersection over Union) ratio between the predicted and ground truth boxes\cite{burns2022motifvln}. Combining multiple methods for a comprehensive evaluation may yield more thorough results.


\subsubsection{\textbf{Multi-dimensional Evaluation}}
\label{sec:5.2.3}
It refers to the evaluation of various abilities of an agent from multiple perspectives. The advent of OSWorld\cite{OSWorld} marks a new phase in the scope and scale of GVA. Researchers need to design evaluation metrics from multiple dimensions to better evaluate these multifaceted systems. We have summarized the following representative multi-dimensional evaluation metrics:

\textbf{Efficiency metric} focuses on efficiency in addition to accuracy. Recent studies collectively assess efficiency through shortest-path navigation~\cite{chen2024webvln}, resource expenditure~\cite{chatdev}, and adaptation metrics such as peak memory allocation and adaptation time~\cite{tack2024onlineadaptationlanguagemodels}. These approaches illustrate diverse methods to quantify an agent's operational efficiency.

\textbf{Human similarity metric} evaluates how closely an agent's behavior mimics human actions. Some works analyze the resemblance between agent operations and human annotations in navigational~\cite{wang2023voyager} and GUI tasks~\cite{lu2024gui}, respectively, emphasizing trajectory and action matching. In conversational settings, Turing Experiment~\cite{turingExp22} assesses the agent's ability to replicate human dialogue patterns. These works contribute to the identification of more human-like agents, providing opportunities for personalized customization.

\textbf{Toxic metric} evaluates an agent's ability to interact without harming users through harmful or biased communication. Lu et al.~\cite{lu2023memochat} and Zheng~\cite{zheng2023secretsrlhflargelanguage} focus on assessing agents' contextual appropriateness and adherence to ethical standards, ensuring responses avoid bias, discrimination, and aggression.

\textbf{Statistical reference metric} aims to evaluate an agent's performance using statistical indicators such as the $F1$ score, which is widely used to evaluate agent's performance as a reference metric. Chen et al.\cite{chen-etal-2021-websrc} calculated $F1$ score to further evaluate the agent's web content Q\&A ability. Sun et al.\cite{burns2022motifvln} evaluated the task interpretability ability of agents based on $F1\,score$ from two categories: feasible and infeasible.

Multi-dimensional evaluation offers a broader perspective on the performance of agents, allowing for more flexible and detailed fine-grained assessments. However, there are many multi-dimensional evaluation efforts but lack unified standards.

\begin{table*}[t]
    \begin{center}
    \captionsetup{font={small,stretch=1.25}, labelfont={bf}}
    \caption{We enumerate both quantitative and qualitative evaluation methods, displaying the names of the representative evaluation matrices for each in a table. We distinguish the matrices with an asterisk (*) where they share the same name but differ in meaning.}  
    \vspace{-5mm}
    \renewcommand{\arraystretch}{1.2}
    \resizebox{1\textwidth}{!}{
        \begin{tabular}{c|cc|c}
        \toprule[1pt]
        
        \textbf{Evaluation Type}                & \multicolumn{2}{c|}{\textbf{Metrics}} & \textbf{References} \\ \hline \hline
        
        \multirow{3}{*}{\textbf{Overall Evaluation}}     & \multicolumn{2}{c|}{$Success\,\,Rate$ / $Task\,\,Success\,\,Rate$} & WebShop\cite{NEURIPS2022_82ad13ec}, MIND2WEB\cite{NEURIPS2023_5950bf29}, WorkArena\cite{workarena2024}, GUI Odyssey\cite{lu2024gui} \\
                                                & \multicolumn{2}{c|}{$Exact\,\,match$ / $exact\_match$ / $Complete\,\,Match$} & WebSRC\cite{chen-etal-2021-websrc}, Webarena\cite{zhou2024webarena}, VisualWebArena\cite{koh2024visualwebarena}, PixelHelp\cite{li-etal-2020-mapping} \\
                                                & \multicolumn{2}{c|}{$Accuracy$} & RUSS\cite{xu-etal-2021-grounding} \\ \hline
                                                                     
        \multirow{9}{*}{\textbf{Detail Evaluation}}      & \multicolumn{1}{c|}{\multirow{3}{*}{Step-wise}} & $Success\,\,Rate$* / $Step\,\,Success\,\,Rate$ & MiniWoB++\cite{ICLR17-Shi}, MIND2WEB\cite{NEURIPS2023_5950bf29}\\ 
                                                & \multicolumn{1}{c|}{} & $Partial\,\,Match$ / $partial\,\,sequence\,\,accuracy$ & PixelHelp\cite{li-etal-2020-mapping}, MoTiF\cite{burns2022motifvln}, AITW\cite{NEURIPS2023_bbbb6308},\\ 
                                                & \multicolumn{1}{c|}{} & $Process\,\,Score$ / $Sequence\,\,Score$ & Mobile-Agent\cite{wang2024mobile}, OmniACT\cite{kapoor2024omniactdatasetbenchmarkenabling} \\ \cline{2-3} 
                                                & \multicolumn{1}{c|}{\multirow{3}{*}{Set Inclusion}} & $Accuracy*$ & WebQA\cite{Chang_2022_CVPR}, META-GUI\cite{sun-etal-2022-meta} \\
                                                & \multicolumn{1}{c|}{} & $must\_include$ & Webarena\cite{zhou2024webarena}, VisualWebArena\cite{koh2024visualwebarena} \\
                                                & \multicolumn{1}{c|}{} & $BLEU\,\,score$ & META-GUI\cite{sun-etal-2022-meta} \\ \cline{2-3}
                                                & \multicolumn{1}{c|}{\multirow{3}{*}{Multi-dimensional}} & $Distance$ & WebVLN\cite{chen2024webvln} \\
                                                & \multicolumn{1}{c|}{} & $Fluency$ & WebQA\cite{Chang_2022_CVPR} \\
                                                & \multicolumn{1}{c|}{} & $F1\,\,score$ / $Operation\,\,F1$ / $average\,\,F1 score$ & WebSRC\cite{chen-etal-2021-websrc}, MIND2WEB\cite{NEURIPS2023_5950bf29}, MoTiF\cite{burns2022motifvln} \\ \hline
        
        \multirow{3}{*}{\textbf{Human Evaluation}}       & \multicolumn{2}{c|}{$Task\,\,Success\,\,Rate$ / $Completion\,\,Rate$} & WebVoyager\cite{he2024webvoyager}, Mobile-Agent\cite{wang2024mobile} \\ 
                                                & \multicolumn{2}{c|}{$Relative\,\,Efficiency$} & Mobile-Agent\cite{wang2024mobile} \\ 
                                                & \multicolumn{2}{c|}{$partial\,\,action\,\,matching\,\,scores$} & AITW\cite{NEURIPS2023_bbbb6308} \\ \hline
                                                                     
        \multirow{3}{*}{\textbf{MLLM-based Evaluation}}  & \multicolumn{2}{c|}{$Accuracy$**} & GUI-WORLD\cite{chen2024guiworld}, AER\cite{pan2024autonomous} \\ 
                                                & \multicolumn{2}{c|}{$fuzzy\_match$} & Webarena\cite{zhou2024webarena} \\
                                                & \multicolumn{2}{c|}{$eval\_vqa$} & VisualWebArena\cite{koh2024visualwebarena} \\ \hline
        \end{tabular}
    }
    \label{tab:Classification_of_metrics}
    \end{center}
    \vspace{-5mm}
\end{table*}

\subsection{\textbf{Human Evaluation}}
\label{sec:5.3}

Human evaluation, relying on \textbf{high-quality human supervision}, is crucial for judging aspects that automated tools cannot accurately assess, such as human likeness in a Turing Test scenario. This method stands as an intuitive and reliable approach for evaluating GVA, providing personalized, qualitative evaluations that offer detailed insights into an agent's ability to mimic human behavior and interaction. It involves trained evaluators who assess a broad array of tasks according to human interaction standards, ensuring evaluations align with end-user experiences. Notable implementations include WebVoyager\cite{he2024webvoyager}, which utilizes binary classification to determine task completion, and AITW\cite{NEURIPS2023_bbbb6308}, assessing step-by-step accuracy in a mobile app context. Additionally, Mobile-Agent\cite{wang2024mobile} introduces the \textit{Relative\,\,Efficiency}, comparing the agent's steps to manual steps to gauge human-like efficiency. These methods underscore the user-centered evaluation crucial for refining GVA capabilities and ensuring their practical usability.

Human evaluation, however, faces challenges such as high costs, low efficiency, and evaluator bias. The manual process increases labor costs and slows down processing. Moreover, different evaluators may introduce cognitive biases, complicating the reproducibility and comparability of results.

\subsection{\textbf{MLLM-based Evaluation}}
\label{sec:5.4}

To address previous limitations, MLLM-based evaluation has become increasingly popular, representing a shift in evaluation methods by replacing human supervisors with advanced models such as GPT-4\cite{achiam2023gpt} and BLIP-2. These efficient evaluations leverage intelligent MLLMs with trillions of parameters that possess powerful understanding capabilities to tackle various issues. Projects like VisualWebArena\cite{koh2024visualwebarena} and WebVoyager\cite{he2024webvoyager} utilize these models to assess similarity and task completion by analyzing agents' outputs against ground truths. Additionally, studies like GUI-WORLD\cite{chen2024guiworld} validate MLLM effectiveness against human evaluation by using MLLMs to score open-ended responses and dialogues, highlighting their capability to efficiently and accurately handle evaluations that traditionally required human oversight. Employing such intelligent models achieves cost-effective, efficient, and convenient assessments.

MLLM-based evaluation, utilizing advanced models like GPT-4, offers efficiency, convenience, and cost-effectiveness. When combined with human evaluation, this approach leverages the best of both worlds: the rapid analytical capabilities of MLLMs and the nuanced understanding of human evaluators. Together, they provide a more thorough and precise assessment method, enhancing evaluation processes across various fields.

\section{Limitations}
\label{sec:6}

Although substantial work has invigorated the agent community, it is undeniable that not all tasks within virtual environments are yet entrusted to agents. In scenarios such as unfamiliar tasks, cross-query operations, and online payments, manual intervention remains prevalent. This reflects three significant limitations in the development of GVAs: insufficient transferability, limited long-sequence decision-making, and heightened security concerns.

\subsection{\textbf{Unrealistic environment and dataset}}
\label{sec:6.1}
It must be acknowledged that, despite its popularity in academia, the agent has not reached the heights anticipated for it. This shortfall is attributable to its operating environments not being sufficiently realistic and the data not being adequately comprehensive—failing to simulate the countless states and scenarios of the real world, and thus unable to train truly generalist agents experts in every field. To address this dilemma, datasets must be expanded and environments enriched to a level of realism surpassing current works represented by operating systems~\cite{OSWorld}, hardware devices~\cite{li-etal-2020-mapping}, and applications~\cite{zhang2023appagentmultimodalagentssmartphone}. At that time, the observational space for GVA would encompass all digital information, necessitating meticulous design and extensive adaptation efforts. This implies a need for prolonged investment and escalated costs. Consequently, long-term planning and careful consideration are essential to navigate these challenges effectively.

\subsection{\textbf{Insufficient Transferability}}
\label{sec:6.2}


Despite proficient performance on Google Maps, the same agent often underperforms on Apple Maps, highlighting discrepancies that may stem from overfitting or differences in platform logic. These variations indicate that GVAs frequently focus on operational procedures without mastering the underlying principles of the tasks they perform. This situation underscores the necessity for grounded learning in the development of universal digital agents. A deeper understanding of a task’s core logic across various platforms is crucial for fostering a truly autonomous generalist agent. Grounded learning not only aids in recognizing the nuances between platforms but also enhances adaptability, crucial for ensuring consistent user experiences across different devices and operating systems. Recent advancements, such as GPT-4v's success in real-time task completion through manually grounded text plans~\cite{zheng2024gpt4visiongeneralistwebagent}, and the proposal for interactive, visually grounded language learning~\cite{suglia2024visually}, further emphasize the benefits of this approach in extending GVA capabilities beyond mere dataset definitions to achieving intrinsic task goals.

\subsection{\textbf{Limited Long-Sequence Decision-making}}
\label{sec:6.3}


In practice, agents often struggle with operations such as ``rollback" due to the lack of an internal world model, which is crucial for predicting environmental conditions and simulating long-term outcomes. This deficiency hinders their ability to plan and adapt as humans do, involving iterative refinement and exploration of different reasoning paths. To improve agents' long-sequence decision-making capabilities, it is crucial to integrate mechanisms that handle both past interactions and future predictions. MemoryBank~\cite{zhong2024memorybank} utilizes a memory update mechanism that mimics human memory retention based on the Ebbinghaus forgetting curve, enhancing recall of past interactions. For future-oriented strategies, the Q*~\cite{wang2024q} model provides a heuristic for estimating expected rewards, and RAP~\cite{hao2023reasoning} employs a large language model coupled with a Monte Carlo tree search to balance exploration and exploitation, thereby enhancing decision-making across temporal and textual spans. Together, these approaches significantly boost the robustness and adaptability of agents.

\subsection{\textbf{Heightened Security Concerns}}
\label{sec:6.4}

The extensive system permissions required for intelligent agents to operate across platforms introduce significant security and privacy risks, as personalized services often necessitate access to substantial amounts of data, exacerbating the conflict between personalization and privacy. To ensure security, one solution is the deployment of edge-based intelligent agents~\cite{zhu2024llava}, involving the lightweight transformation of large models for full implementation on mobile devices~\cite{xiao2023smoothquant}—a significant test of model lightweight. Alternatively, edge-cloud collaborative technologies~\cite{lin2024awq} enable intelligent agents to process data on endpoint devices and, when necessary, coordinate efficiently with cloud services. This approach reduces the centralized storage and sensitive data transmission, enhancing both privacy and security.

\section{Future}
\label{sec:7}

\subsection{\textbf{From Individual to Systematic}}
\label{sec:7.1}
The domain of AI is experiencing a significant paradigm shift, moving from the deployment of isolated agents toward the adoption of systematic agent frameworks~\cite{han2024llm}. In this context, the term ``systematic'' does not merely refer to an increase in the number of agents, but rather to developing a dedicated layer that seamlessly integrates all operations between users and operating systems. This layer delegates complex and redundant low-level processes to specialized agents~\cite{luzolo2024combining},~\cite{shah2024multi}, much like how operating systems revolutionized the coordination and allocation of software and hardware resources. What this agent system ultimately presents to the user is a highly personalized and user-friendly interaction experience.

The current trajectory of the computing industry clearly reflects this trend: on content platforms, agents assist users in generating and modifying content; in operating systems, Apple has introduced Apple Intelligence, which facilitates the foundational integration of agent capabilities at the system level. In hardware, various tensor processing units (TPUs) are designed to support the computational demands of agent-based models. These developments strongly indicate the evolving role of agent systems, suggesting that in the future, we may witness more intelligent agents being integrated as an independent system layer, further transforming the way we interact with technology.

\subsection{\textbf{From Virtual to Physical}}
\label{sec:7.2}
As we move beyond the confines of purely software-based systems, the incorporation of physical embodiment opens new avenues for enhancing the adaptability and functionality of intelligent systems. This shift is not merely about adding physical components to existing algorithms; it's about rethinking intelligence as an emergent property of the dynamic interplay between an agent and its environment. Embodied agents, equipped with sensors and actuators, are not just passive recipients of pre-programmed instructions but are active participants in their environment. This enables them to learn from real-world interactions and refine their behaviors in ways that virtual agents, confined to digital realms, cannot. The result is a more robust form of intelligence that can better understand and respond to the complexities of real-world scenarios, paving the way for autonomous robotics, personalized healthcare, and interactive entertainment systems.

Embodied intelligence, as an emerging research perspective and methodology, is reshaping our understanding of the nature and development of agents. While virtual agent-based intelligence often views intelligence as an abstract capability independent of physical presence, embodied intelligence posits that intelligence arises from continuous interaction between an entity and its environment~\cite{xu2024survey}. The core of embodied intelligence computing systems provides foundational capabilities in perception, cognition, and action for various forms of robots~\cite{varley2024embodied}, enabling developers to build more complex and diverse robotic applications. Through such systems, robots can achieve a deeper understanding of and more effective interaction with the physical world, resulting in more natural and efficient service and collaboration.

\section{Conclusion}
\label{sec:8}
In this survey, we review the Generalist Virtual Agent (GVA), a new archetype that uses multimodal data to autonomously operate within virtual environments. Our systematic examination of existing research reveals that GVAs are more likely to exhibit human-like intelligence when they operate in environments that mimic the real world. However, there is a significant over-reliance on large-scale models, which poses a risk should their development plateau. Thus, we propose that GVAs evolve from traditional tools into sophisticated agent systems that enhance human-computer interaction, and we advocate for a breakthrough into embodied intelligence that extends beyond digital confines. This approach could significantly expand the capabilities and applications of GVAs, potentially transforming their role in technology and society.

\bibliographystyle{ieeetr}
\bibliography{Reference}

\begin{IEEEbiography}[{\includegraphics[width=1in,height=1.25in,clip,keepaspectratio]{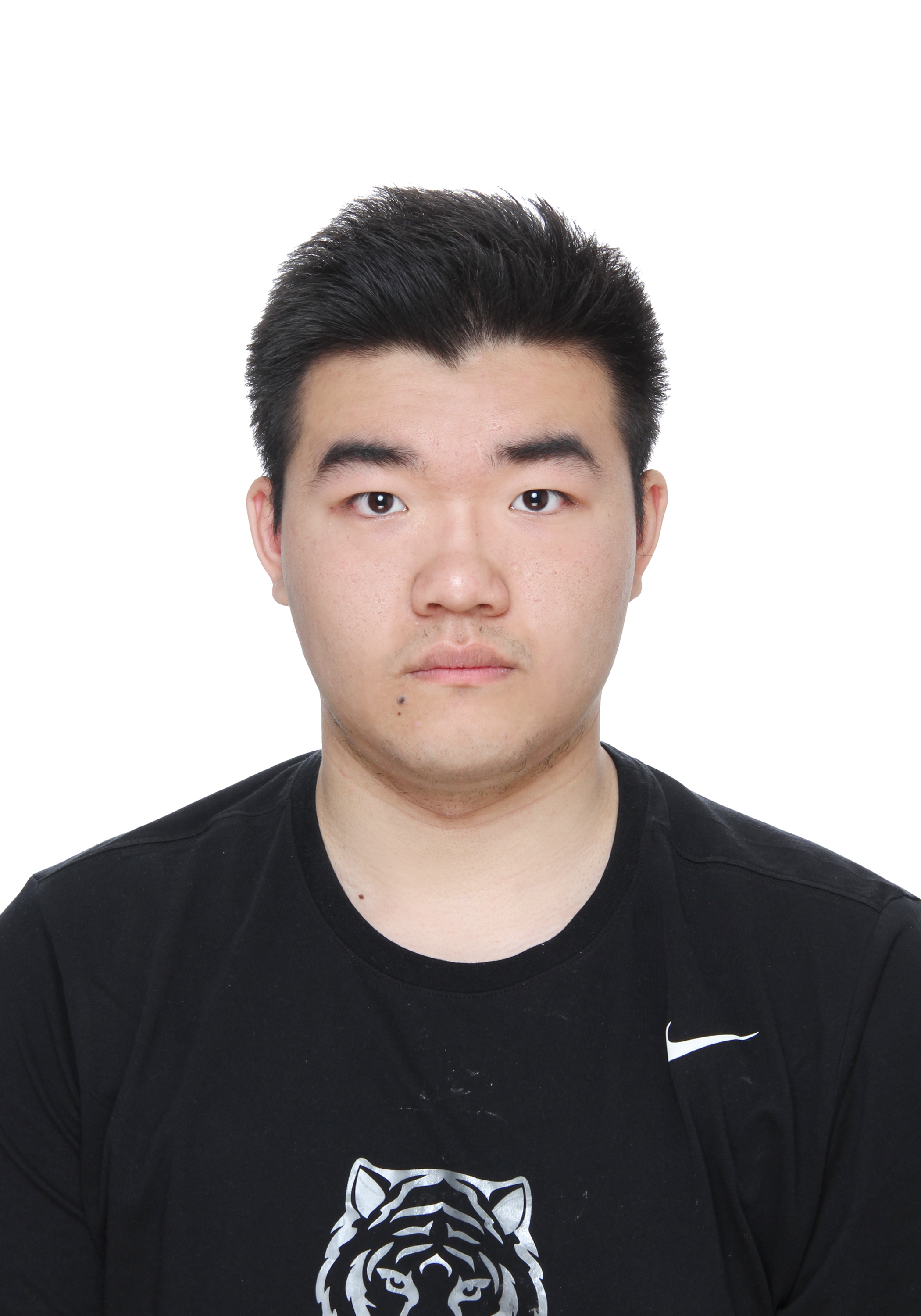}}]
	{Minghe Gao} is currently a PhD student at Zhejiang University. He previously obtained his bachelor degree from the Chu Kochen Honors College at Zhejiang University. His research interests include the training and application of multimodal large language models and agents based on multimodal models. He is a recipient of the Zhejiang Provincial Outstanding Graduate Award. So far, he has published papers at ICCV, ICLR, and MM.
\end{IEEEbiography}

\begin{IEEEbiography}[{\includegraphics[width=1in,height=1.25in,clip,keepaspectratio]{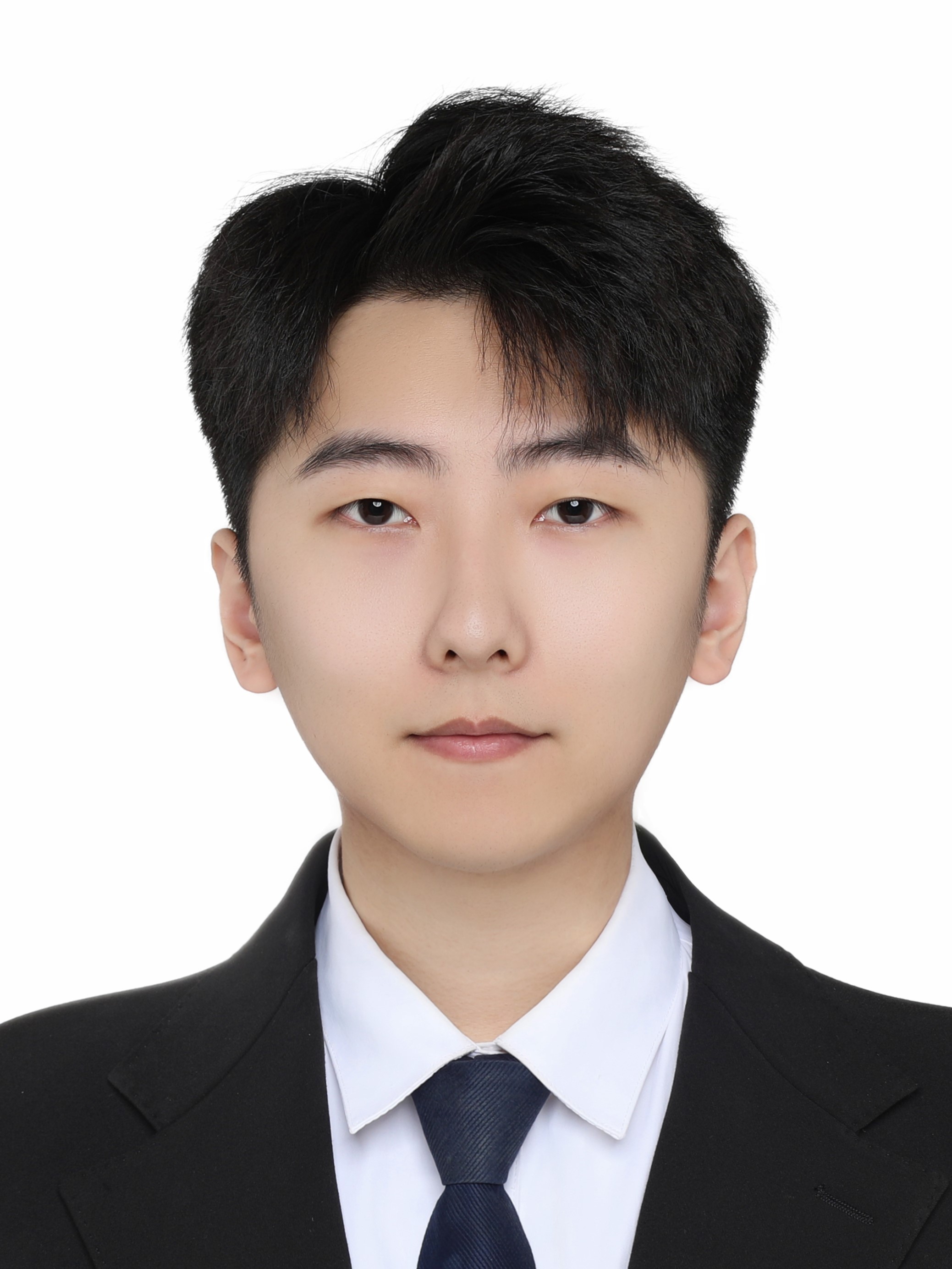}}]
	{Wendong Bu} is currently a master student at Zhejiang University. His research interests include the application of multimodal large language models in the field of agents. He is a recipient of the Shandong Provincial Outstanding Graduate Award.
\end{IEEEbiography}

\begin{IEEEbiography}[{\includegraphics[width=1in,height=1.25in,clip,keepaspectratio,angle=90]{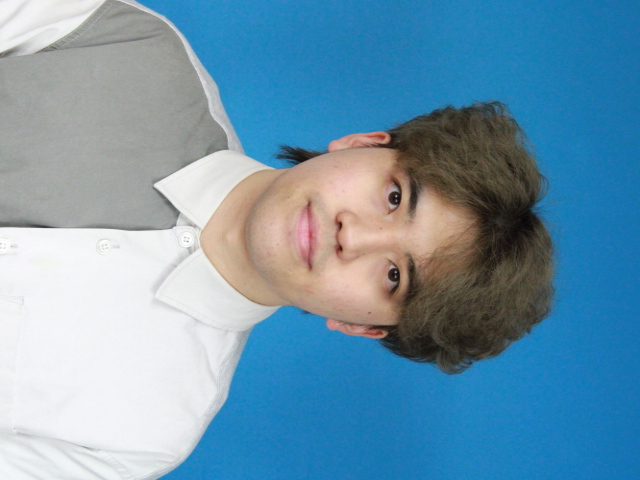}}]
	{Bingchen Miao} is currently a master's student at Zhejiang University. He previously obtained his undergraduate degree from Central China Normal University. His research interests include multimodal learning and agents based on multimodal models.
\end{IEEEbiography}

\begin{IEEEbiography}[{\includegraphics[width=1in,height=1.25in,clip,keepaspectratio]{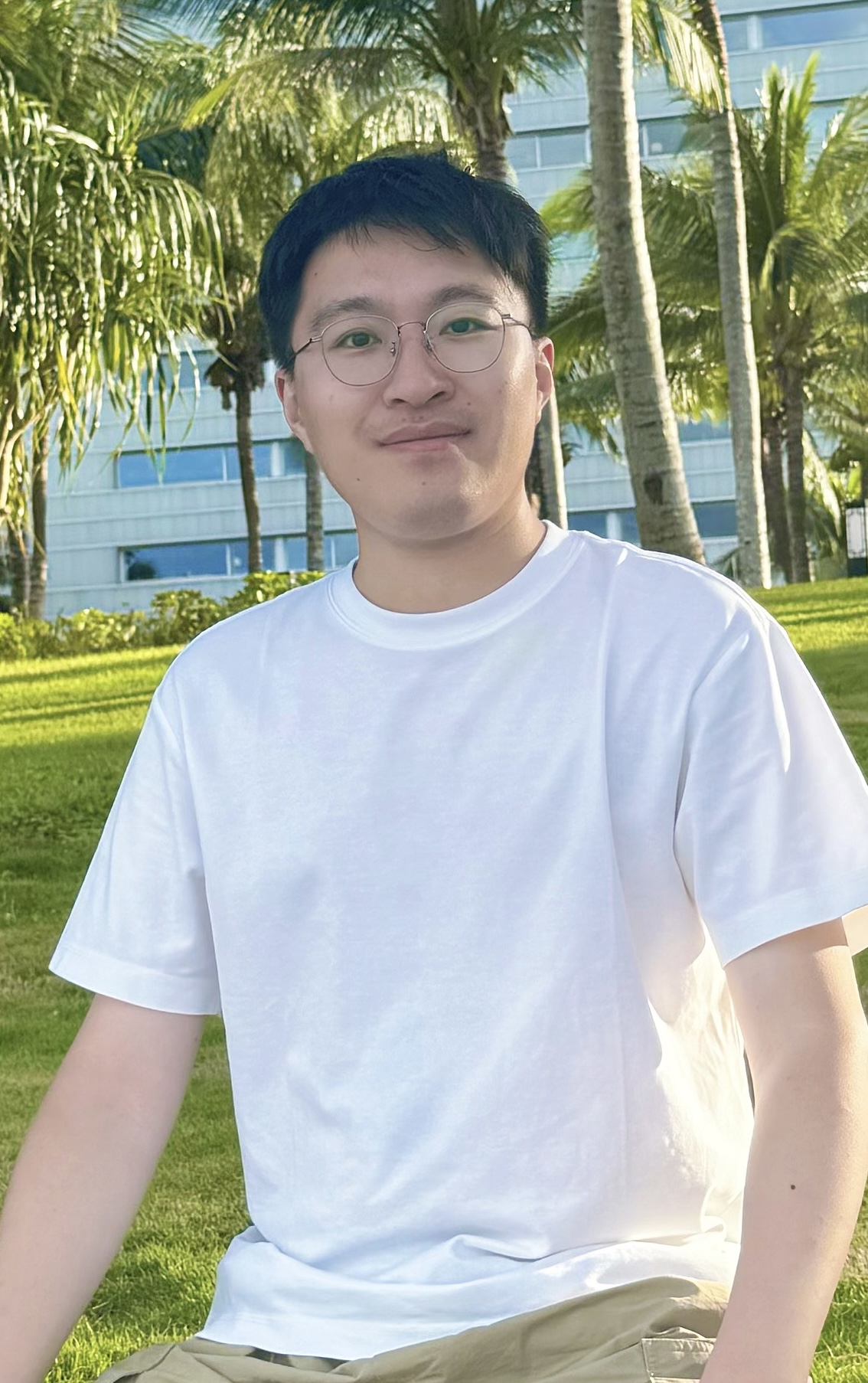}}]
	{Yang Wu} is currently a senior algorithm engineer at Alipay.com Co., Ltd. He received the BEng degree from Southeast University, China, and the MSc degree in computer science from the University of Edinburgh, U.K. His research interests include large language models, agents, and information retrieval. 
\end{IEEEbiography}

\begin{IEEEbiography}[{\includegraphics[width=1in,height=1.25in,clip,keepaspectratio]{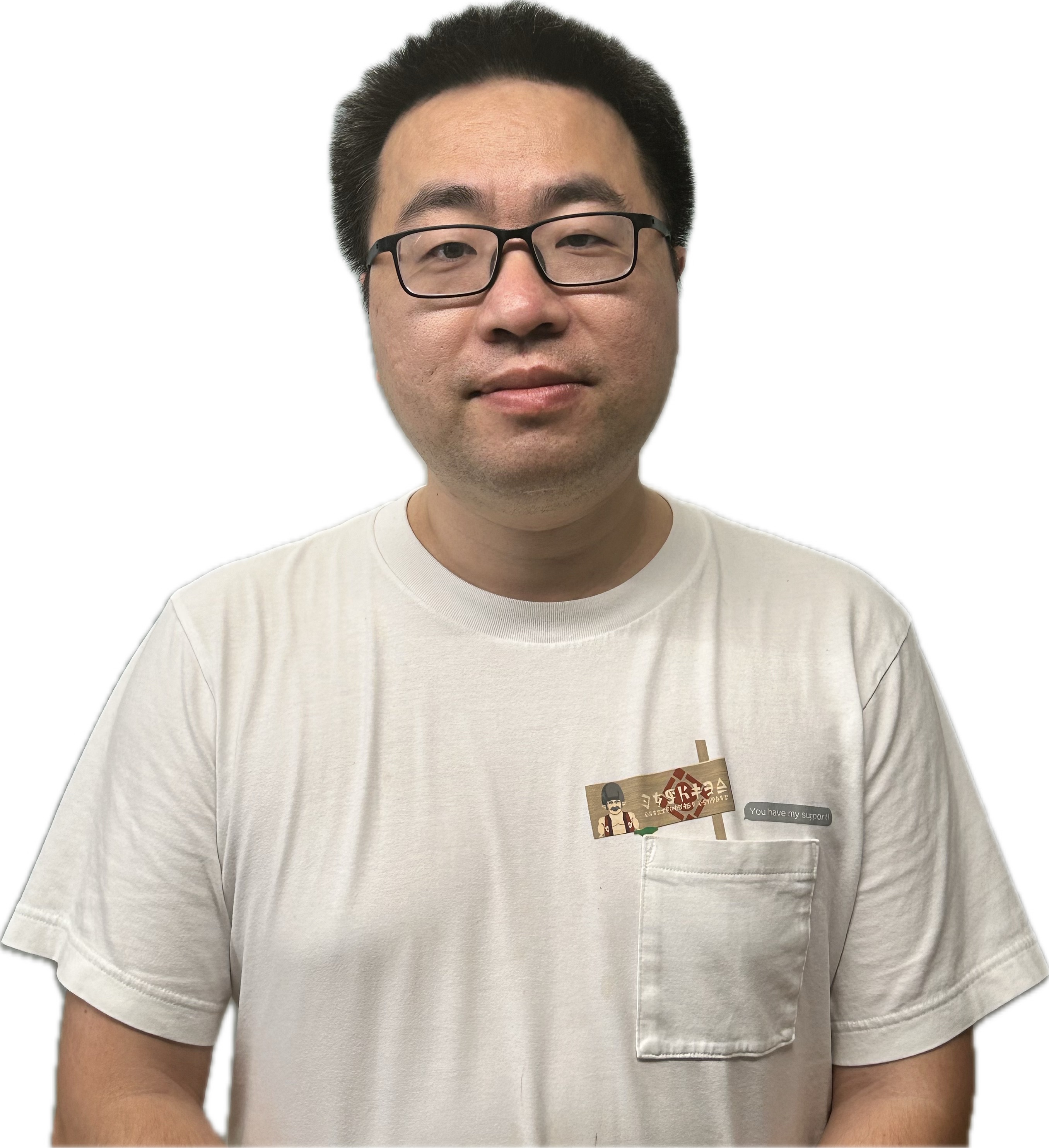}}]
	{Yunfei Li} received the BSc and MSc degree both in computer science from Zhejiang University, Hangzhou, China. He is currently a principle algorithm engineer at  Alipay.com Co., Ltd., his research interests include recommendation systems and large language models.
\end{IEEEbiography}

\begin{IEEEbiography}[{\includegraphics[width=1in,height=1.25in,clip,keepaspectratio]{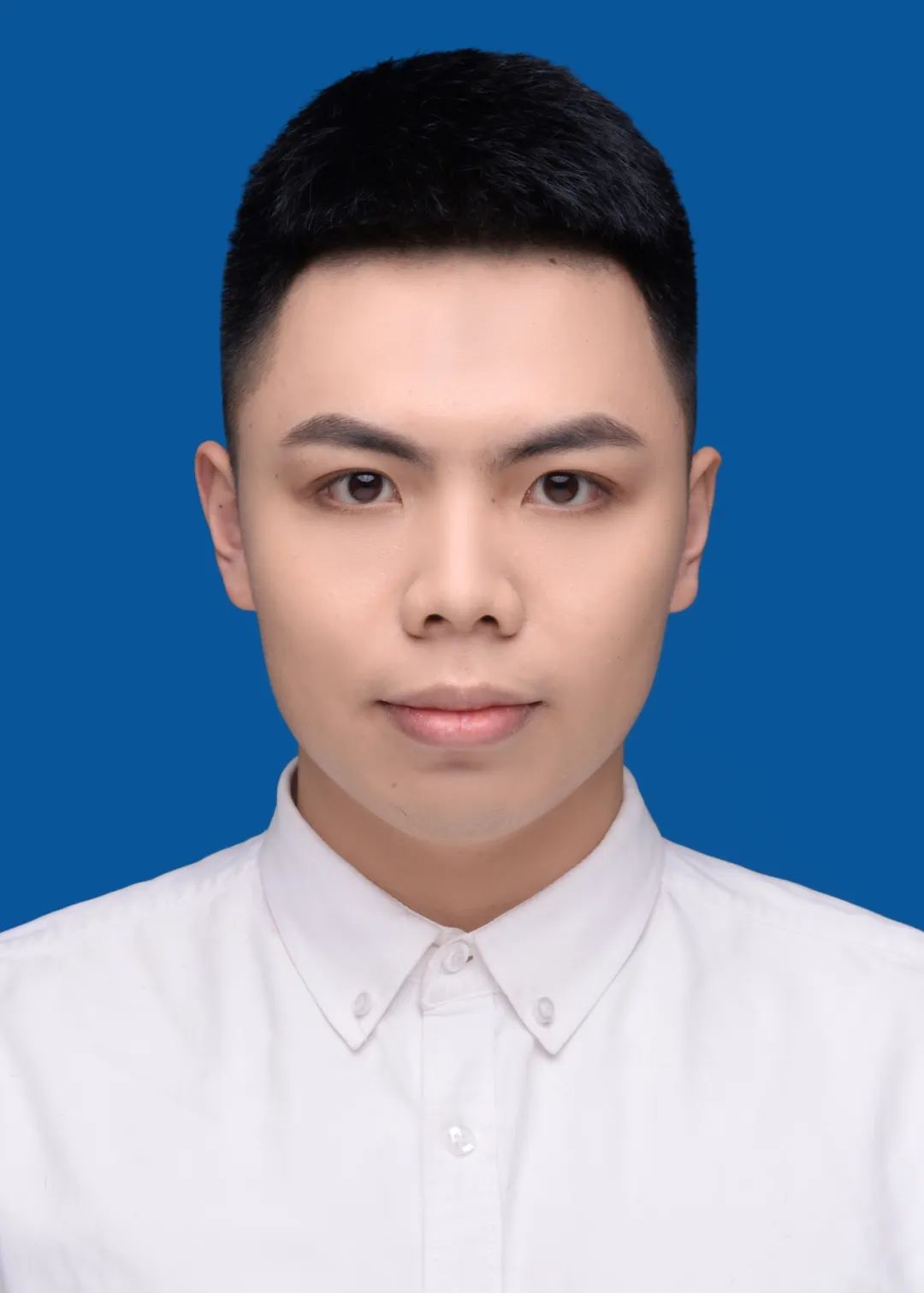}}]
	{Juncheng Li} is currently a ZJU100 Young Professor at Zhejiang University. He received his Ph.D. degree from the Zhejiang University, Hangzhou, China, in 2023. His research interests include multimodal learning and video understanding. So far, he has published more than 30 papers in top conferences such as CVPR, ICCV, NeurIPS, ICLR, ICML, ACM MM, SIGIR, and journals including TPAMI, and TKDE.	
\end{IEEEbiography}

\begin{IEEEbiography}[{\includegraphics[width=1in,height=1.25in,clip,keepaspectratio]{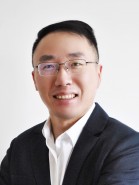}}]
	{Siliang Tang} is currently a full professor at the College of Computer Science, Zhejiang University. In 2012, Siliang got his Ph.D. degree at the National University of Ireland, Maynooth. His research interests include multimodal analysis, information extraction and etc. So far, he has published more than 100 papers in top-tier scientific conferences such as AAAI, IJCAI, NIPS, ICML, KDD, CVPR, ICCV, ACL, SIGMM, SIGIR, and IEEE journals such as IEEE TKDE, IEEE TIP, IEEE TVCG, IEEE TMM, IEEE TCSVT, IEEE TNNLS. He has been serving as an area chair/senior PC member or PC member in conferences such as NIPS, ICML, KDD, AAAI, IJCAI, CVPR, ACL, EMNLP, NAACL, SIGMM and reviewers of journals such as IEEE TIP, IEEE TMM, IEEE TSMC, ACM Computing Surveys, Nature- Scientific Reports, etc. He is a member of the IEEE.
\end{IEEEbiography}

\begin{IEEEbiography}[{\includegraphics[width=1in,height=1.25in,clip,keepaspectratio]{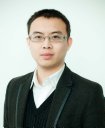}}]
	{Qi Wu} received the MSc and PhD degrees in computer science from the University of Bath, U.K., in 2011 and 2015, respectively. He is currently an Associate Professor and an ARC DECRA Fellow at the University of Adelaide. He won the Australian Academy of Science J G Russell Award in 2019. He is also a CI of the Australian Centre for Robotic Vision (ACRV) and a Program Leader of the Australian Institute of Machine Learning (AIML). Before that, he worked as a Postdoc Researcher in the Australian Centre for Visual Technologies (ACVT). He received an MSc in Global Computing and Media Technology, a PhD in Computer Science from the University of Bath (United Kingdom), in 2011 and 2015. His work has been published in prestigious journals and conferences such as IEEE TPAMI, CVPR, ICCV, ECCV and AAAI. 
\end{IEEEbiography}

\begin{IEEEbiography}[{\includegraphics[width=1in,height=1.25in,clip,keepaspectratio]{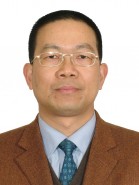}}]
	{Yueting Zhuang} (Senior Member, IEEE) received the B.Sc., M.Sc., and Ph.D. degrees in computer science from Zhejiang University, China, in 1986, 1989, and 1998, respectively. From February 1997 to August 1998, he was a Visiting Scholar with the University of Illinois at Urbana-Champaign. From 2009 to 2017, he served as the Dean of the College of Computer Science, Zhejiang University, and the Director of the Institute of Artificial Intelligence from 2006 to 2015. He is now a CAAI Fellow (2018) and serves as the standing committee member of CAAI. He is a Fellow of China Society of Image and Graphics (2019). Also he is a member of Zhejiang Provincial Government AI Development Committee (AI Top 30). Currently, he is a Full Professor with the College of Computer Science and the Director of the MOE-Digital Library Engineering Research Center, Zhejiang University. His research interests include multimedia retrieval, video-language understanding, cross-media computing, and digital library. He has won various awards and honors such as National Science Fund for Distinguished Young Scholars of China from National Natural Science Foundation(2005), the "Chang Jiang Scholars Program" Professor of the Ministry of Education of China (2008), the chief scientist of 973 Project (2012CB316400). He was the leading PI of the digital library project— “China America Digital Academic Library (CADAL)”, which has now become one of the largest non-profit digital libraries in the world. Also he is the director of the technical center of UNESCO Category II International Knowledge Center of Engineering Science and technology (IKCEST). He has published over 700 papers with 23,000 citations by Google Scholar.
\end{IEEEbiography}

\begin{IEEEbiography}[{\includegraphics[width=1in,height=1.25in,clip,keepaspectratio]{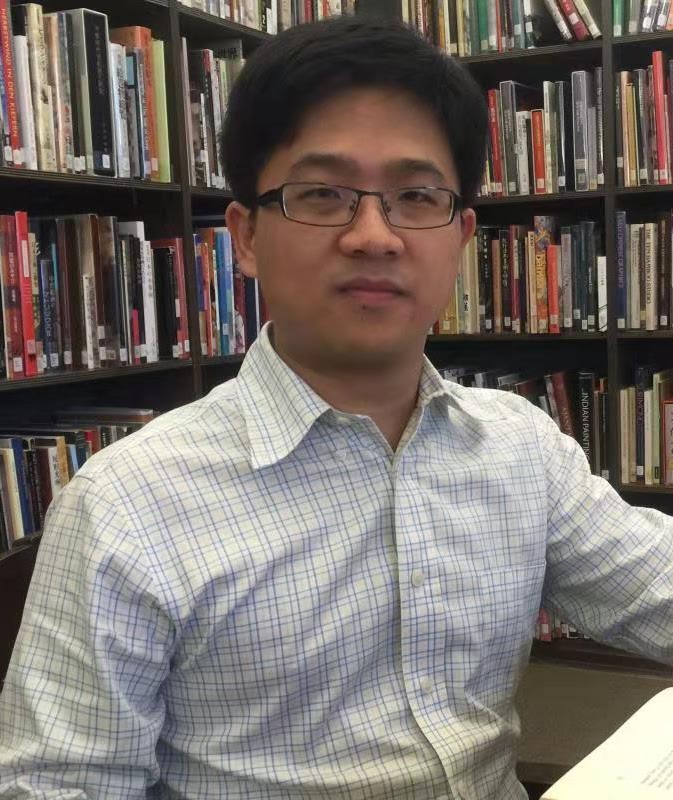}}]
	{Meng Wang} is a professor at Hefei University of Technology, China. He received the B.E. degree and Ph.D. degree in the Special Class for the Gifted Young and the Department of Electronic Engineering and Information Science from the University of Science and Technology of China (USTC), Hefei, China, in 2003 and 2008, respectively.He worked as an associate researcher at Microsoft Research Asia and a senior research fellow at National University of Singapore. His current research interests include multimedia content analysis, computer vision, and pattern recognition. He has authored or co-authored over 200 book chapters, journal and conference papers. He holds over 30 US, Chinese, and international granted patents. He received paper prizes or awards from ACM MM 2009 (Best Paper Award), ACM MM 2010 (Best Paper Award), MMM 2010 (Best Paper Award), ICIMCS 2012 (Best Paper Award), ACM MM 2012 (Best Demo Award), ICDM 2014 (Best Student Paper Award), SIGIR 2015 (Best Paper Honorable Mention), IEEE TMM 2015 and 2016 (Prize Paper Award Honorable Mention), IEEE SMC 2017 (Best Transactions Paper Award), and ACM TOMM 2018 (Nicolas D. Georganas Best Paper Award). He is or has been an editorial board member of IEEE Trans. on Circuits and Systems for Video Technology, IEEE Trans. on Multimedia, IEEE Trans. on Knowledge and Data Engineering, IEEE Trans. on Neural Networks and Learning Systems, etc. He is the General Co-Chair of ICMR 2021, PCM 2018 and MMM 2013, and the Program Co-Chair of ICIMCS 2013.
\end{IEEEbiography}

\end{document}